\documentclass{class_file_sig}

\usepackage{epsfig}    
\usepackage[colorlinks,citecolor=black,filecolor=black,linkcolor=black,urlcolor=black]{hyperref}
\usepackage{amsmath}
\usepackage{dsfont}
\usepackage{verbatim}
\usepackage{slashbox}
\usepackage{multirow} 
\usepackage{rotating} 
\usepackage{ctable}
\usepackage{slashbox}

\usepackage{array}
\usepackage{makecell}

\usepackage[labelformat=simple]{subfig}

\newcommand{\etal}{\emph{et~al.}}

\usepackage[normalem]{ulem} 

\makeatletter
\def\tagform@#1{\maketag@@@{\ignorespaces#1\unskip\@@italiccorr}}
\let\orgtheequation\theequation
\def\theequation{(\orgtheequation)}
\makeatother

\makeatletter
\let\@copyrightspace\relax
\makeatother

\begin{document}
\pagenumbering{arabic}
\sloppy
\title{Network Characteristics of Video Streaming Traffic}
\numberofauthors{3}
\author{
\alignauthor Ashwin Rao\\
\affaddr{INRIA France}
\alignauthor Yeon-sup Lim\\
\affaddr{University of Massachusetts, Amherst, MA}
\alignauthor Chadi Barakat\\
\affaddr{INRIA France}
\and
\alignauthor Arnaud Legout\\
\affaddr{INRIA France}
\alignauthor Don Towsley\\
\affaddr{University of Massachusetts, Amherst, MA}
\alignauthor Walid Dabbous\\
\affaddr{INRIA France}
}
\maketitle

\begin{abstract}

Video streaming represents a large fraction of Internet
traffic. Surprisingly, little is known about the network
characteristics of this traffic. In this paper, we study the network
characteristics of the two most popular video streaming services,
Netflix and YouTube. We show that the streaming strategies vary
with the type of the application (Web browser or native mobile
application), and the type of container (Silverlight, Flash, or HTML5)
used for video streaming. In particular, we identify three different
streaming strategies that produce traffic patterns from non-ack
clocked ON-OFF cycles to bulk TCP transfer. We then present an
analytical model to study the potential impact of these streaming
strategies on the aggregate traffic and make recommendations
accordingly. 
\end{abstract}
\keywords{Video streaming\footnote{\em This is the author version of the
    paper accepted for publication at ACM CoNEXT 2011, December 6--9
    2011.}, Streaming strategies, YouTube, Netflix,
  Silverlight, HTML5, Flash.} 

\section{Introduction}
\label{sec:introduction}

The popularity of video streaming has considerably increased in the
last decade. Indeed, recent studies have shown that video streaming is
responsible for 25-40\% of all Internet
traffic~\cite{Sandvine_Internet2011,Maier_2009_ResBroadband}. The two
dominant sources for video streaming traffic in North America are
Netflix and YouTube~\cite{Sandvine_Internet2011}. YouTube is also the
most popular source of video streaming traffic in Europe and Latin
America~\cite{Sandvine_Internet2011,Maier_2009_ResBroadband}.

Despite this popularity, little is known about the strategies used by
YouTube and Netflix to stream their videos. These 
strategies might have a fundamental impact on the network traffic. TCP
is used to transport this traffic, but if this traffic is rate
controlled by the application, and this rate is lower than the
end-to-end available bandwidth, the traffic characteristics will not
be the one of a standard TCP flow. This might have an impact
on the network and the traffic coming from other applications. 
In addition, most of the streaming sessions are interrupted due to
lack of interest~\cite{Finamore_2011_YouTubeDeviceInfra,
Gill_2007_Youtube, Huang_2007_VODProfit}. Because of this, the
streaming strategies may have a significant impact on the network
traffic. Indeed, the amount of video downloaded but not watched is an
overhead for the network.
 
In this paper, we present an in depth network traffic analysis of
YouTube and Netflix. In particular, we consider the impact of the
application (Web browsers and the applications for mobile devices),
and the container (Flash~\cite{AdobeFlash},
HTML5~\cite{Hickson_2011_Html5}, Silverlight~\cite{MSSilverlight}), on
the characteristics of the traffic between the source and the
viewer. Then we present a mathematical model to evaluate the impact of
the streaming strategy on the aggregate data rate of video streaming
traffic.  

\noindent
Our contributions are the following:

1)We identify three different streaming strategies with
fundamentally different traffic properties ranging from bulk TCP file
transfer to non-ack clocked traffic.

2) We detail the network traffic characteristics of the three 
streaming strategies currently used by YouTube and Netflix. 

3)We show that the streaming strategy depends on the application and
the container used to stream videos. Therefore, the increased adoption
of one could have a significant impact on the network traffic
characteristics. For instance, following a massive adoption of HTML5
instead of Flash, or an increase in the usage of mobile applications. 

4) We derive a mathematical model to evaluate the impact of the
streaming strategies on the stochastic properties of the aggregate video
streaming traffic. Our model can be used to dimension the network
for video streaming. In particular, it sheds light on the importance
of the different video streaming parameters for traffic
engineering. For example, we show that an increase in the video
encoding rates shall produce smoother aggregate video streaming
traffic. We also present the video streaming parameters that can be
adapted to minimize the amount of unused bytes on user interruptions
due to lack of interest. 
 
The remainder of the paper is organized as follows. In
\autoref{sec:Background} we provide an overview of video streaming. We
then present the three different streaming strategies we identified in
\autoref{sec:streamingStrats}. We discuss the datasets and measurement
techniques in \autoref{sec:Methodology}. We detail the network
characteristics of the streaming strategies used by YouTube and
Netflix in \autoref{sec:MeasurementResults}. In \autoref{sec:Model} we
present our model and discuss the potential impact of these streaming
strategies on the aggregate video streaming traffic. We discuss the
related work in \autoref{sec:RelatedWork} and present our conclusions
in \autoref{sec:Conclusion}.  

\section{Video Streaming Background}
\label{sec:Background}

Video streaming enables viewers to start video playback while the
content is being downloaded. The two dominant sources for video
streaming traffic in the Internet are Netflix and
YouTube~\cite{Sandvine_Internet2011}. Users can view Netflix and
YouTube videos either on PCs, using a Web browser, or on mobile
devices, using a Web browser or a mobile application. A mobile
application is the native Netflix or YouTube application running on
mobile devices. In this paper, for the mobile devices, we exclusively
consider the native YouTube and Netflix applications for the iOS and
Android devices.   

YouTube, one of the most popular sites for user generated videos,
supports two containers for video streaming, Adobe
Flash~\cite{AdobeFlash} and HTML5~\cite{Hickson_2011_Html5}. Adobe
Flash, henceforth referred to as Flash, is the default container when
YouTube is accessed via a PC. Users need to install a proprietary
plugin for viewing Flash videos. HTML5 supports videos that do not
require any proprietary plugins. HTML5 is the default container when
YouTube videos are streamed using the native mobile application for
Android and iOS.  Recently, YouTube has started supporting \emph{High
  Definition} (HD) streaming. The default container for HD videos is
Flash.  

Netflix uses Microsoft Silverlight~\cite{MSSilverlight} to stream
videos. As of today, Netflix does not support any other
containers for video streaming even though Netflix is leveraging
HTML5 for streaming. While streaming to a Web browser requires a
Silverlight plugin, the mobile devices require the native Netflix
application.  

Netflix and YouTube use TCP to stream videos. During a typical
streaming session, apart from the video content, the streaming servers
send other auxiliary data. For example, the auxiliary data includes
details of related videos and advertisements. In this paper, we
restrict ourselves to the TCP connections that are used to transfer
the video content. We are interested in these TCP connections because
these connections contribute to the bulk of the traffic generated by
video streaming.

\section{Streaming Strategies}
\label{sec:streamingStrats}

\begin{figure}
\centering
\includegraphics[width=0.75\columnwidth]{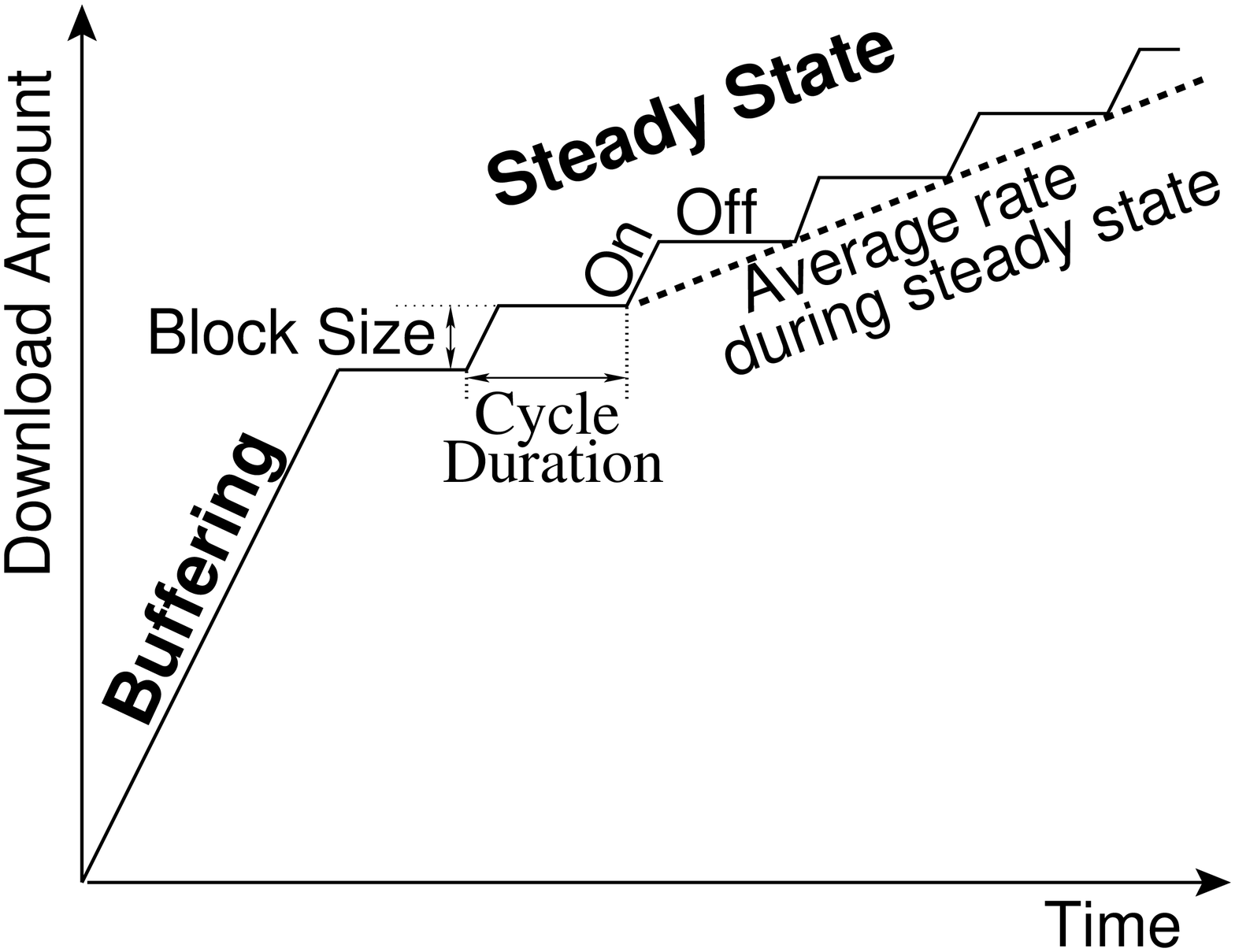}
\caption{Phases of video download. \emph{Video streaming begins
    with a buffering phase followed by a steady state phase. Cycles of
    ON-OFF periods in the steady state phase are used to limit the download
    rate.}}     
\label{fig:StreamingTrafficPatterns}
\end{figure}
In this section, we present the three different streaming strategies
that we identified using the experiments described in
\autoref{sec:MeasurementResults}. Our goal here is to
synthesize the main characteristics of those strategies and present
some of their advantages and disadvantages.

During a typical streaming session, the video content is transferred
in two phases: a \emph{buffering phase} followed by a \emph{steady
state phase}. During the buffering phase, the data transfer rate is
limited by the end-to-end available bandwidth. In
\autoref{fig:StreamingTrafficPatterns}, the slope of the line during
the buffering phase is the end-to-end available bandwidth. The video
player begins playback when a sufficient amount of data is 
available in its buffer. Video playback does not wait for the
buffering phase to end.

In the steady state phase, the average download rate is slightly
larger than video encoding rate. We call the ratio of the average
download rate during the steady state phase and the video encoding
rate the \emph{accumulation ratio}. An accumulation ratio of at least
one is desirable because an accumulation ratio lower than one can
cause the video playback to interrupt due to empty buffers. An
accumulation ratio larger than one implies that the amount of video
content present in the players buffer increases during the steady
state phase, which improves the resilience to transient network
congestion.

The average download rate in the steady state phase is achieved by
periodically transferring one block of video content. These periodic
transfers produce cycles of ON-OFF periods. During each ON period, a
block of data is transferred at the end-to-end available bandwidth
that can be used by TCP; the TCP connection is idle during the OFF
periods. The slope of the download amount during the ON periods in
\autoref{fig:StreamingTrafficPatterns} represents the end-to-end
available bandwidth. We call the amount of data transferred in one
cycle the \emph{block size}.

The buffering phase ensures that the player has a sufficient amount of
data to compensate for the variance in the end-to-end available
bandwidth during video playback. The reduced transfer rate in
the steady state phase ensures that the amount of video content does
not overwhelm the video player while keeping the amount of buffered
data during the buffering phase constant or increasing. The reduced
data transfer rate is important for mobile devices which may not be
able to store the entire video. We also believe that the reduced rate
during the steady state phase reduces the load on the streaming
infrastructure. The reduced load can increase the number of videos
that can be streamed in parallel. 

We use the existence of the steady state phase and the technique used
to throttle the data transfer rate in the steady state phase to
identify the underlying streaming strategy. We observe the following
three streaming strategies for Netflix and YouTube videos.

1) \emph{No ON-OFF Cycles}. For this streaming strategy, all data
is transferred during the buffering phase. As a consequence, we do not
observe a steady state phase for this streaming strategy. An advantage
of this strategy is that it requires no complex engineering at the
server and the client. The video streaming session can be considered
as a simple file transfer session. One disadvantage of this strategy
is that it can overwhelm the player and cause a large amount of unused
bytes if users interrupt the video playback. 

2) \emph{Short ON-OFF cycles}. We define this streaming strategy as
the periodic transfer of blocks of size less than 2.5~MB (called an ON
period) followed by an idle period (called the OFF period). The goal
of this streaming strategy is to maintain an accumulation ratio which
is slightly larger than one. This is achieved by a periodic 
transfer of a block of data followed by an OFF period. An OFF period
is observed only when the average data transfer rate is smaller
than the end-to-end available bandwidth. We do not observe OFF
periods, and short ON-OFF cycles, when the end-to-end available
bandwidth is less than or equal to the average data transfer
rate. This strategy ensures that the client is not overwhelmed by the
amount of data sent by the server.  

3) \emph{Long ON-OFF cycles}. This streaming strategy produces a
traffic pattern that resembles the periodic execution of buffering
phases following long idle periods. The primary difference between
this strategy and the strategy of short ON-OFF cycles is the amount of
data transferred in a cycle. The amount of data transferred during the
ON periods for this strategy is larger than 2.5~MB. For a given
average rate during the steady state phase, the cycle duration for the
strategy of long ON-OFF cycles is longer than the cycle duration for
the strategy of short ON-OFF cycles. This streaming strategy is a
hybrid of the no ON-OFF cycles and short ON-OFF cycles streaming
strategies.

\section{Methodology}
\label{sec:Methodology}

We now present the six datasets used in our measurements and the
technique used to capture the TCP packets while streaming the videos
in each dataset.

\subsection{Dataset}
\label{sec:Dataset}

We first created four datasets of YouTube videos and two datasets of
Netflix videos. The \emph{YouFlash}, \emph{YouHtml}, \emph{YouHD}, and
\emph{YouMob} dataset contain YouTube videos while the \emph{NetPC}
and the \emph{NetMob} dataset contain Netflix videos.

For the \emph{YouFlash}, \emph{YouHD}, \emph{YouHtml}, and
\emph{YouMob} dataset, we respectively searched for Flash videos, HD
videos, HTML5 videos, and videos that can be played by the native iOS
and Android application.  The \emph{YouFlash} and \emph{YouHD}
datasets respectively contain randomly selected 5000 Flash videos and
2000 HD videos. The \emph{YouHtml} dataset contains 2500 videos from
the \emph{YouFlash} dataset and 500 videos from the \emph{YouHD}
dataset; these videos can be played using the HTML5 player. For the
\emph{YouMob} dataset, we searched for videos using the native YouTube
application on an iPad.

The videos in the \emph{YouFlash} and \emph{YouHD} datasets have
encoding rates from 0.2~Mbps to 1.5~Mbps, and 0.2~Mbps to 4.8~Mbps
respectively. The videos in the \emph{YouFlash} dataset have a default
resolution of either 240p or 360p while videos in the \emph{YouHD}
dataset have a default resolution of 720p. The videos in the
\emph{YouFlash} and \emph{YouHD} dataset are streamed using Flash as
the default container. The encoding rate of videos in the
\emph{YouHtml} and \emph{YouMob} dataset is from 0.2~Mbps to 2.5~Mbps,
0.2~Mbps to 2.7~Mbps respectively. When using the HTML5 container to
stream videos to PCs, YouTube uses 360p as the default resolution;
users need to manually switch to a higher resolution such as 720p for
viewing the video in HD. As it is currently not possible to view HD
videos using HTML5 on PCs without manual intervention, we believe that
the fraction of users viewing HD videos using HTML5 on PCs will be
small. In this paper, for PCs, we restrict our study to Flash videos
played at the default resolution, HD videos streamed using Flash, and
HTML5 videos streamed at the default resolution of 360p. We use the
default setting because
Finamore~\etal~\cite{Finamore_2011_YouTubeDeviceInfra} observed that
users use the default player configuration while streaming YouTube
videos. We henceforth refer to videos in the \emph{YouFlash} dataset
as Flash videos, videos in the \emph{YouHD} dataset as HD videos, and
videos in the \emph{YouHtml} dataset as HTML5 videos.

For Netflix datasets, we collected the list of 11208 videos available
for watching instantly as of 20-May-2011. Then, we randomly selected
200 videos from this list for the \emph{NetPC} dataset. For the
\emph{NetMob} dataset we randomly selected 50 videos from \emph{NetPC}
dataset.

\subsection{Measurement Technique}
\label{sec:MeasurementTechnique}

We now present the list of software tools used for our
measurements. We used Internet Explorer 9~\cite{IE9}, Mozilla Firefox 
4.0~\cite{Firefox}, and Google Chrome 10.0~\cite{GoogleChrome}
(henceforth referred to as Chrome) for streaming videos on PCs. These
three browsers have a combined usage share of more than
80\%~\cite{BrowserShare}. For Flash videos, we installed the Flash
plugin 10.2 in each of these browsers. For Netflix videos, we
installed Microsoft Silverlight 4.0.60310. For HTML5 videos, we
installed the webM codec in Internet Explorer as YouTube uses
webM~\cite{WebM} as the default codec used for HTML5 videos. Firefox
and Chrome have a built-in support from webM. We used
tcpdump~\cite{TCPDUMP} on Linux and windump~\cite{WINDUMP} on Windows
to capture the packets exchanged between the web browser and the
streaming servers. To study the streaming strategies used for mobile
applications, we used an Android (version 2.2) smart-phone and an iPad (iOS
version 4.2.1). We used the native YouTube and Netflix applications,
developed by YouTube and Netflix respectively for these mobile
devices.

We captured the packets exchanged during video streaming in the
following manner. When a PC was used for streaming videos, we serially
iterated through the list of videos in each dataset and performed the
following steps for each video. We first started tcpdump, or windump
depending on the operating system, to capture the packets
exchanged. We then started a web browser and loaded the URL of a video
on the same machine to start the video streaming session. We stopped
the streaming session and the packet capture after 180 seconds. For
native mobile applications we first started the packet capture on a
machine that can access the packets exchanged between the mobile
application and the streaming server. We then started the video
streaming. We stopped the packet capture and streaming after 180
seconds. 

We performed our measurements from the following four locations.

1) A 100 Mbps wired connection connected to the Internet through a
  500 Mbps link. We refer to this network as the \emph{Research 
    network} in the rest of the paper.    

2) A 54 Mbps Wi-fi connection behind a ADSL router with typical
  download rate of 7.7 Mbps and an upload rate of 1.2 Mbps. This
  network is referred to as the \emph{Residence network} in the rest
  of the paper.

3) A 100~Mbps wired connection connected to the Internet through a
  1~Gbps link. We refer to this network as the \emph{Academic network}
  in the rest of the paper.

4) A 100Mbps wired connection behind a cable modem connected to the 
  Comcast ISP; we observe a typical download rate of 20Mbps and an
  upload rate of 3Mbps in this network. We refer to this network as
  the \emph{Home network} in the rest of the paper. 

The Research and the Residence networks are based in France, while the
Academic and the Home networks are based in the United States of
America. The YouTube measurements were carried out from each of these
four locations. The Netflix measurements were carried out only in the
Academic and Home networks because Netflix currently does not stream
videos to France.  For the native mobile applications, the YouTube
measurements were carried out in the \emph{Research} network by using
a 54~Mbps Wi-fi connection; the Netflix measurements were carried out
using a 54~Mbps Wi-fi connection in the \emph{Academic} network. The
YouTube measurements were carried out from 01-Feb-2011 to
30-May-2011. The Netflix measurements were carried out from
20-May-2011 to 14-Jun-2011. 

\section{Measurement Results}
\label{sec:MeasurementResults}

The goal of this section is to present an in depth analysis of YouTube
and Netflix traffic and to show that the video streaming traffic
generated by YouTube and Netflix can be classified in the three
streaming strategies discussed in \autoref{sec:streamingStrats}.   

\begin{table}
\begin{footnotesize}
\begin{tabular}{|>{\centering} m{0.20\columnwidth} |
    >{\centering}m{0.12\columnwidth} |
    >{\centering}m{0.15\columnwidth} |
    >{\centering}m{0.12\columnwidth} |
    >{\centering}m{0.18\columnwidth}|}
\hline
Service & \multicolumn{3}{c|}{\bf YouTube} & {\bf Netflix}
\tabularnewline
\hline
Container & Flash & HTML5 & Flash HD & Silverlight
\tabularnewline
\hline
Internet Explorer & Short & Short & No & Short 
\tabularnewline
\hline
Mozilla Firefox & Short & No & No & Short 
\tabularnewline
\hline
Google Chrome & Short & Long & No & Short 
\tabularnewline
\hline
iOS (native) & \multirow{2}[4]{0.18\columnwidth}{Not Applica-ble} & Multiple &
\multirow{2}[4]{0.18\columnwidth}{Not Applica-ble} & Short 
\tabularnewline
\cline{1-1}
\cline{3-3}
\cline{5-5}
Android (native) &  & Long &  & Long 
\tabularnewline
\hline 
\end{tabular}
\end{footnotesize}
\caption{Streaming Strategies. \emph{Short, Long, and No
     respectively stand for the strategies of short ON-OFF cycles, long
     ON-OFF cycles, and no ON-OFF cycles. Streaming strategy depends on
     the combination of browser and container}.}
\label{tab:StrategyYoutubeNetflix}
\end{table}

\autoref{tab:StrategyYoutubeNetflix} summarizes our finding on the
strategies used to stream Netflix and YouTube videos. While using the
Flash container, we observed that the applications do not throttle the
rate of data transfer; rate control if any is performed by the YouTube
servers. Therefore, in \autoref{tab:StrategyYoutubeNetflix}, the
streaming strategy is independent of the application used for Flash
videos and HD videos. For HTML5 videos, we observed that the YouTube
servers do not explicitly control the data transfer rate. Because,
the applications use their own techniques to throttle the data
transfer rate, we observe that the streaming strategies for HTML5
videos depend on the application used. We observed that Netflix uses
short ON-OFF cycles for streaming videos to PCs irrespective of the
web browser. However, the streaming strategies differ for the native
mobile applications; we observe short ON-OFF cycles for iPad and long
ON-OFF cycles for Android.

To characterize the traffic we use three different metrics: the amount
downloaded during the buffering phase, the blocks size, and the
accumulation ratio. To compute the accumulation ratio we need the
encoding rate of the videos. For videos using the Flash container, we
obtain the video encoding rate from the header of the video file being
streamed. For HTML5 videos, YouTube uses webM as the default codec.
During our measurements, we were unable to determine the encoding rate
of HTML5 videos because we observed an invalid entry for the frame
rate in the header of the webM files. Therefore, we estimate the
encoding rate of HTML5 videos by dividing the {\tt Content-Length}
present in the HTTP response by the duration of the video. For Netflix
videos, we do not use the accumulation ratio because the encoding rate
used by Netflix depends on the end-to-end available
bandwidth~\cite{Akhshabi_2011_StreamingRateAdaption}. 

In this section, we first detail the streaming strategies used by
YouTube and Netflix in \autoref{sec:YouTube} and \autoref{sec:Netflix}
respectively. We then discuss the implications of these strategies in
\autoref{sec:SummaryYouTubeNetflix}.  

\subsection{YouTube Streaming Strategies}
\label{sec:YouTube}

We now detail the buffering phase and the steady state phase of the
three streaming strategies used by YouTube.  

\subsubsection{Short ON-OFF cycles}
\label{sec:YouTubeShort}

\begin{figure}
  \subfloat[Short ON-OFF cycles.]{\label{fig:ShortONOFFDownEvol}\includegraphics[width=0.48\columnwidth]{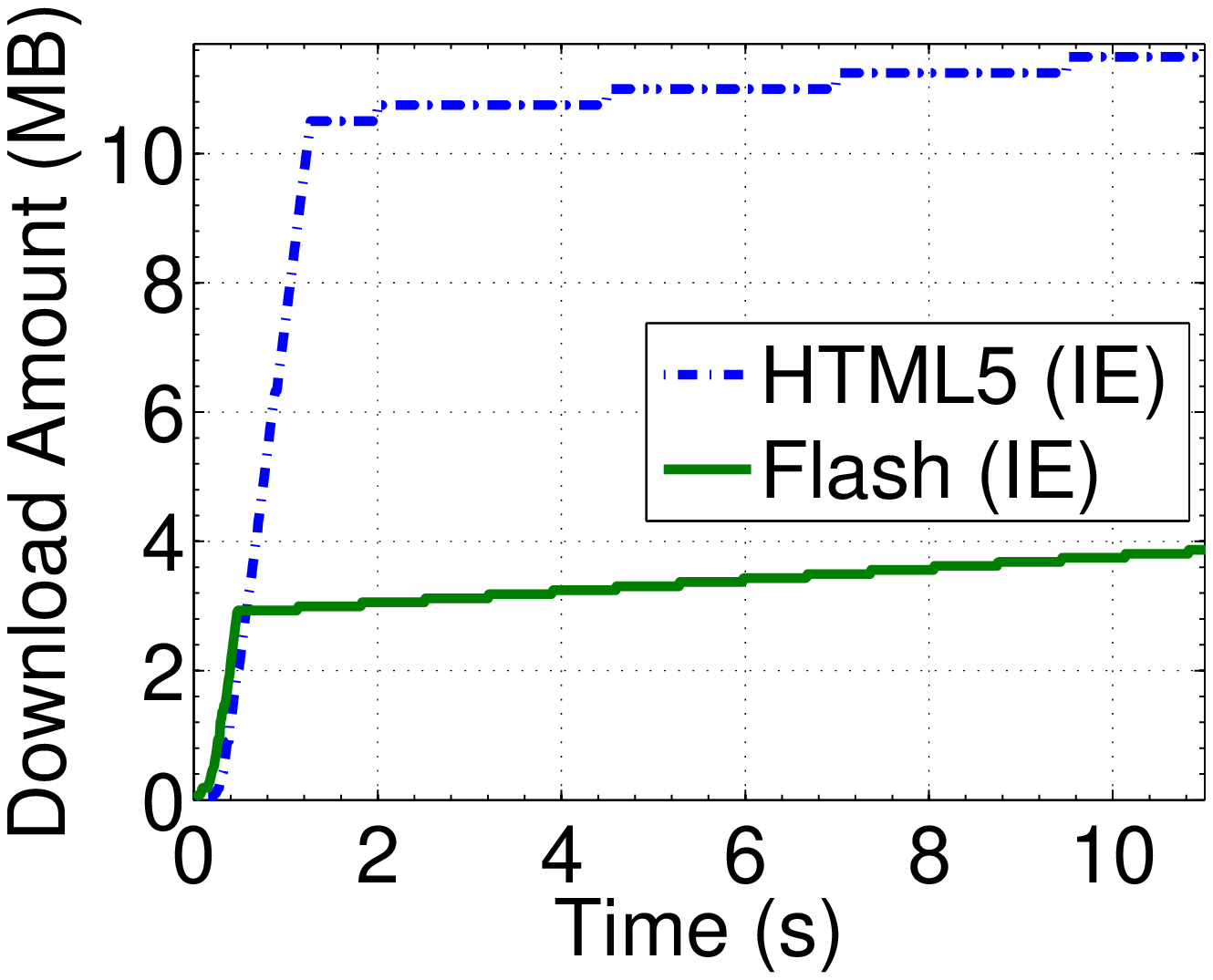}}
  \subfloat[TCP Receive Window.]{\label{fig:ShortONOFFWindowEvol}\includegraphics[width=0.48\columnwidth]{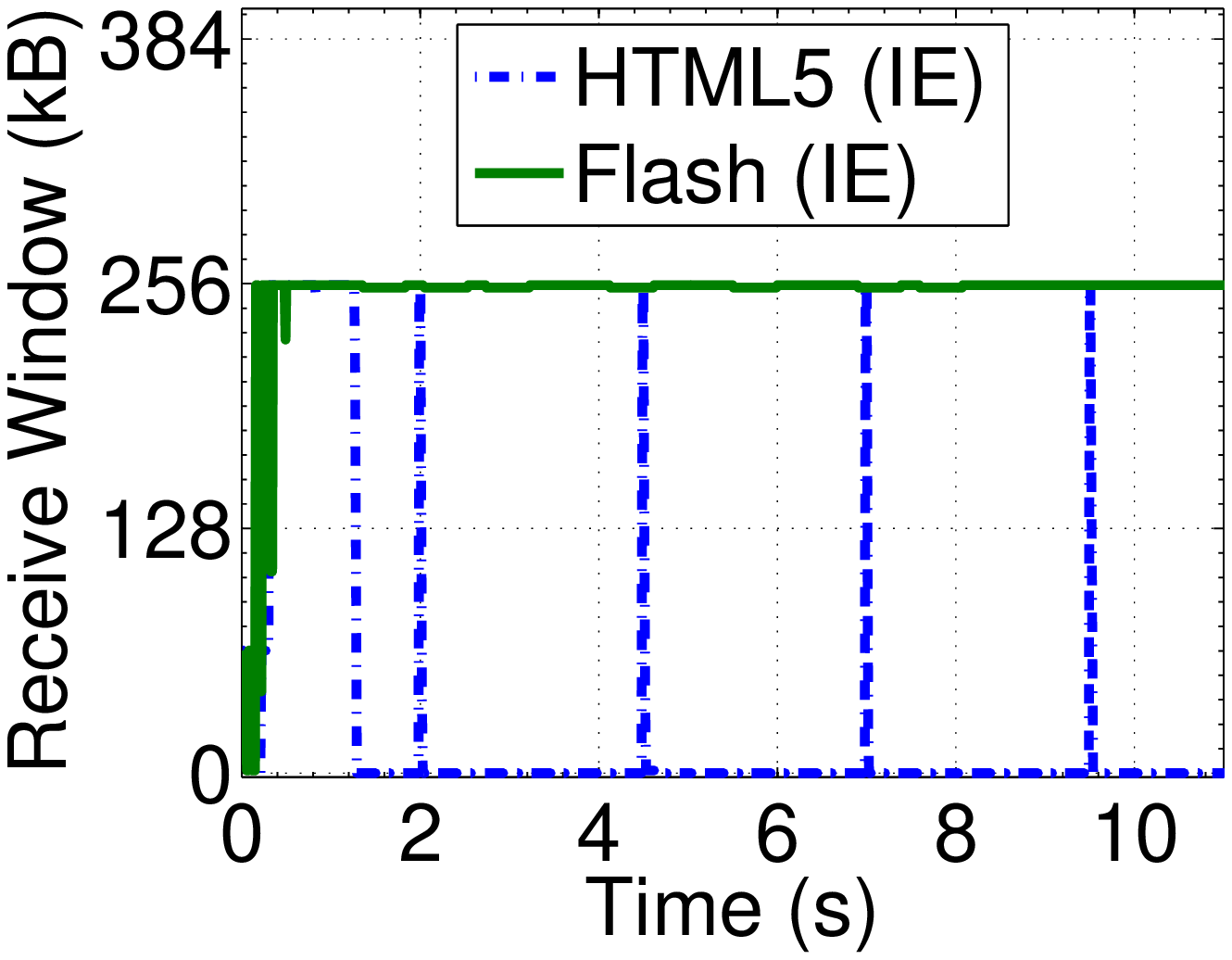}}
  \caption{Short ON-OFF cycles. {\em The evolution of the TCP
      receive window shows that YouTube servers explicitly limit the
      download rate of  Flash videos.}}
  \label{fig:ShortONOFF}
\end{figure}

We observe short ON-OFF cycles for Flash videos regardless of the
browser used, and for HTML5 videos when Internet Explorer is used.

In \autoref{fig:ShortONOFF}, we present a representative
trace observed while streaming one Flash video and one HTML5 video;
the videos were streamed using Internet Explorer (IE) in the Research 
network. For both the videos, in \autoref{fig:ShortONOFFDownEvol}, we
observe a buffering phase followed by a steady state phase. During the
steady state phase the download amount increments in short steps. We
present the evolution of the TCP receive window for the two streaming 
sessions in \autoref{fig:ShortONOFFWindowEvol}. In
this figure we observe that the TCP receive window
periodically becomes empty when streaming the HTML5 video. This
implies that Internet Explorer throttles the download rate of the
HTML5 video by periodically pulling data from the TCP buffers. In
\autoref{fig:ShortONOFFWindowEvol}, we do not observe such explicit
rate control by Internet Explorer when streaming the Flash video. This
implies that, for the Flash video, the YouTube servers throttle the
rate of data transfer by periodically pushing the video content. We
observe this behavior for Flash videos regardless of the
browser. We do not present the supporting figures due to
space constraints.     

We now detail the buffering phase and the steady state phase when
YouTube videos are streamed using the strategy of short ON-OFF
cycles. We use the videos in the \emph{YouFlash} and \emph{YouHtml}
dataset for these measurements.  

\begin{figure}
  \centering
   \subfloat[Flash Video.]{\label{fig:BufferingShortONOFFFlash}\includegraphics[width=0.48\columnwidth]{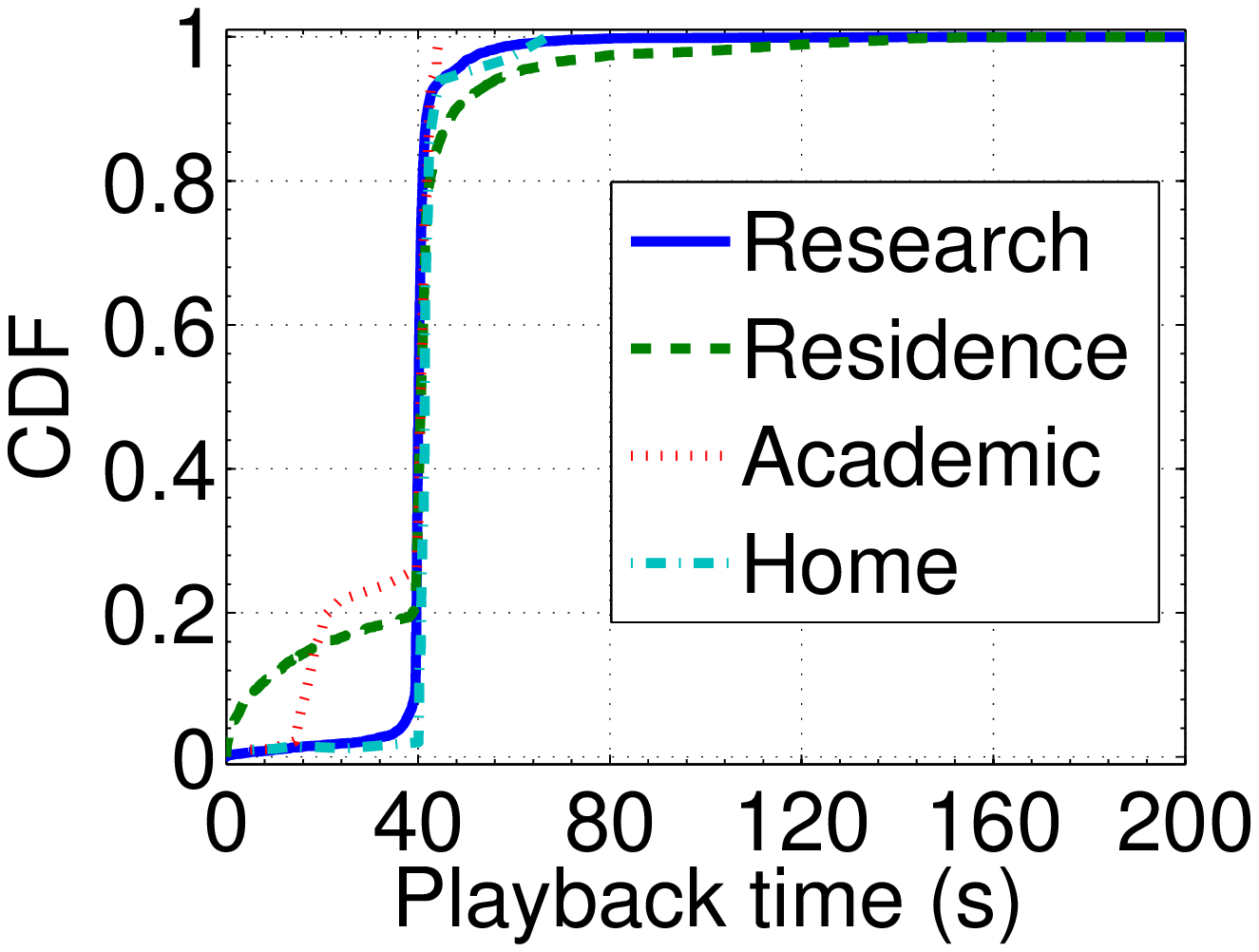}}
  \hspace{0.01\columnwidth}
  \subfloat[HTML5 on Internet Explorer.]{\label{fig:BufferingShortONOFFIEHtml5}\includegraphics[width=0.48\columnwidth]{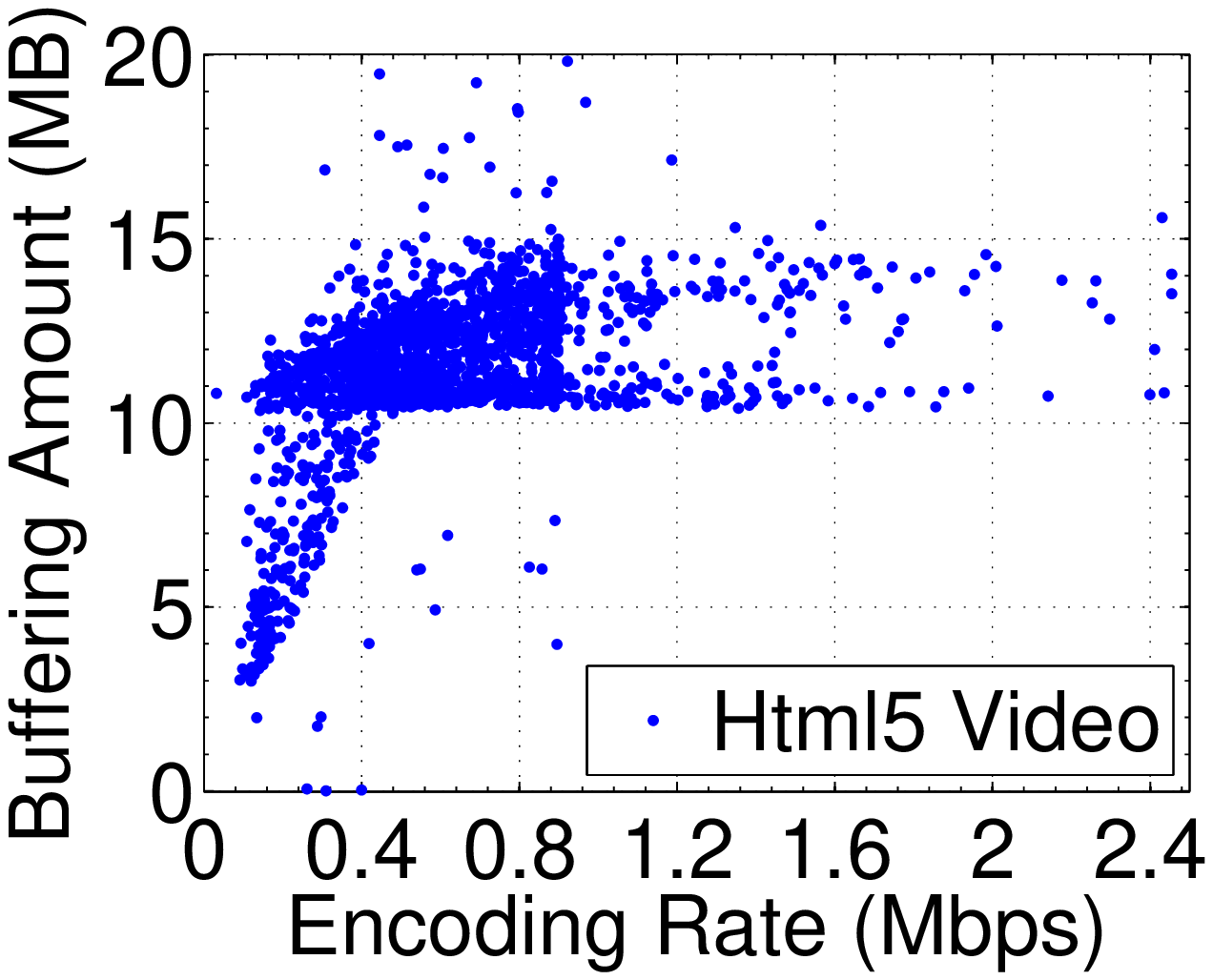}}\\
  \caption{Amount downloaded during the  buffering phase. \emph{For Flash
      videos, approximately  40~seconds worth of
      playback is downloaded in the buffering phase. The buffering amount and
      the video encoding rate is weakly correlated for HTML5
      videos}.}  
\end{figure}

\noindent
\textbf{i) Buffering Phase.}
In \autoref{fig:BufferingShortONOFFFlash} we observe that for most of
the videos in the \emph{YouFlash} dataset, YouTube sends approximately
40~seconds worth of playback data during the buffering phase. The
playback time is calculated by dividing the amount downloaded during
the buffering phase by the video encoding rate. We present the
cumulative distribution (CDF) of the playback time in
\autoref{fig:BufferingShortONOFFFlash}. The steep slope for
the distribution of the playback time is because of the strong
correlation (correlation coefficient = 0.85) between the video
encoding rate and the amount downloaded during the buffering phase.  

For the Residence and the Academic networks, in
\autoref{fig:BufferingShortONOFFFlash}, we observe a smaller amount of
buffering. The smaller amount could be an artifact of our technique
used to measure the amount downloaded during the buffering phase;
we consider the start time of the first OFF period as the end of the
buffering phase. This technique is sensitive to packet losses and we
observed higher packet retransmissions, median of 1.02\% and 0.76\%
respectively, in the Residence network and the Academic network.  

For HTML5 videos, in \autoref{fig:BufferingShortONOFFIEHtml5} we
observe that the amount of data downloaded during the buffering phase is
not strongly correlated to the video encoding rate (correlation
coefficient = 0.41). The results presented in
\autoref{fig:BufferingShortONOFFIEHtml5} are for the Research
Network. We make similar observations for other networks. 

In summary, we observe that the YouTube servers push 40~seconds of
playback data during the buffering phase for Flash videos. For HTML5
videos, Internet Explorer typically downloads from 10~MB to 15~MB
during the buffering phase. Therefore, the buffering phase for
HTML5 videos streamed to Internet Explorer can be more aggressive. For
example, for a video encoding rate of 1~Mbps, 10~MB corresponds to 
80~seconds of playback time.

\label{sec:ShortONOFFSteady}
\begin{figure}
  \centering
   \subfloat[Block size.]{\label{fig:BlockSizeShortONOFFFlash}\includegraphics[width=0.48\columnwidth]{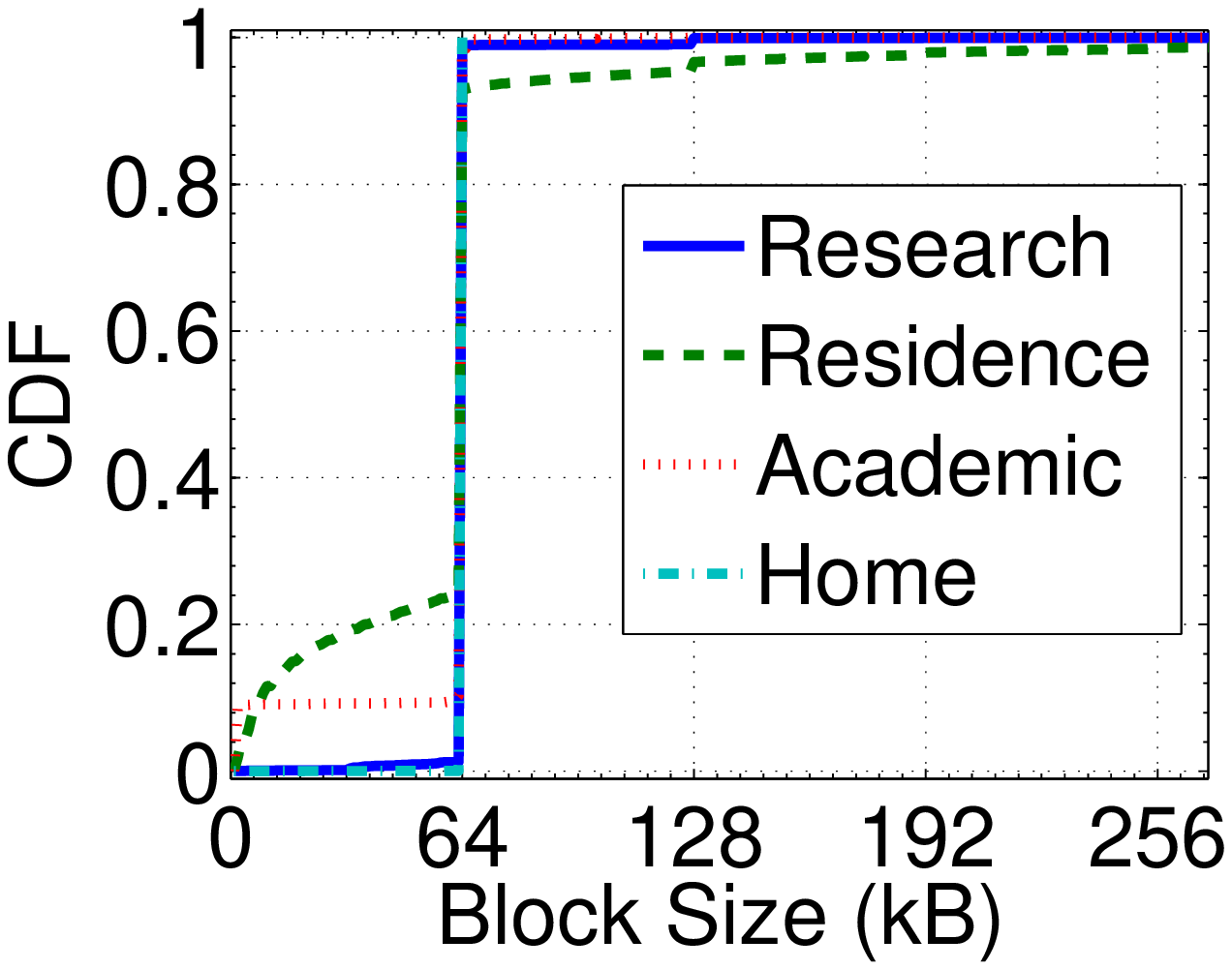}}
  \hspace{0.01\columnwidth}
  \subfloat[Accumulation Ratio.]{\label{fig:AccumulationRatioShortONOFFFlash}\includegraphics[width=0.48\columnwidth]{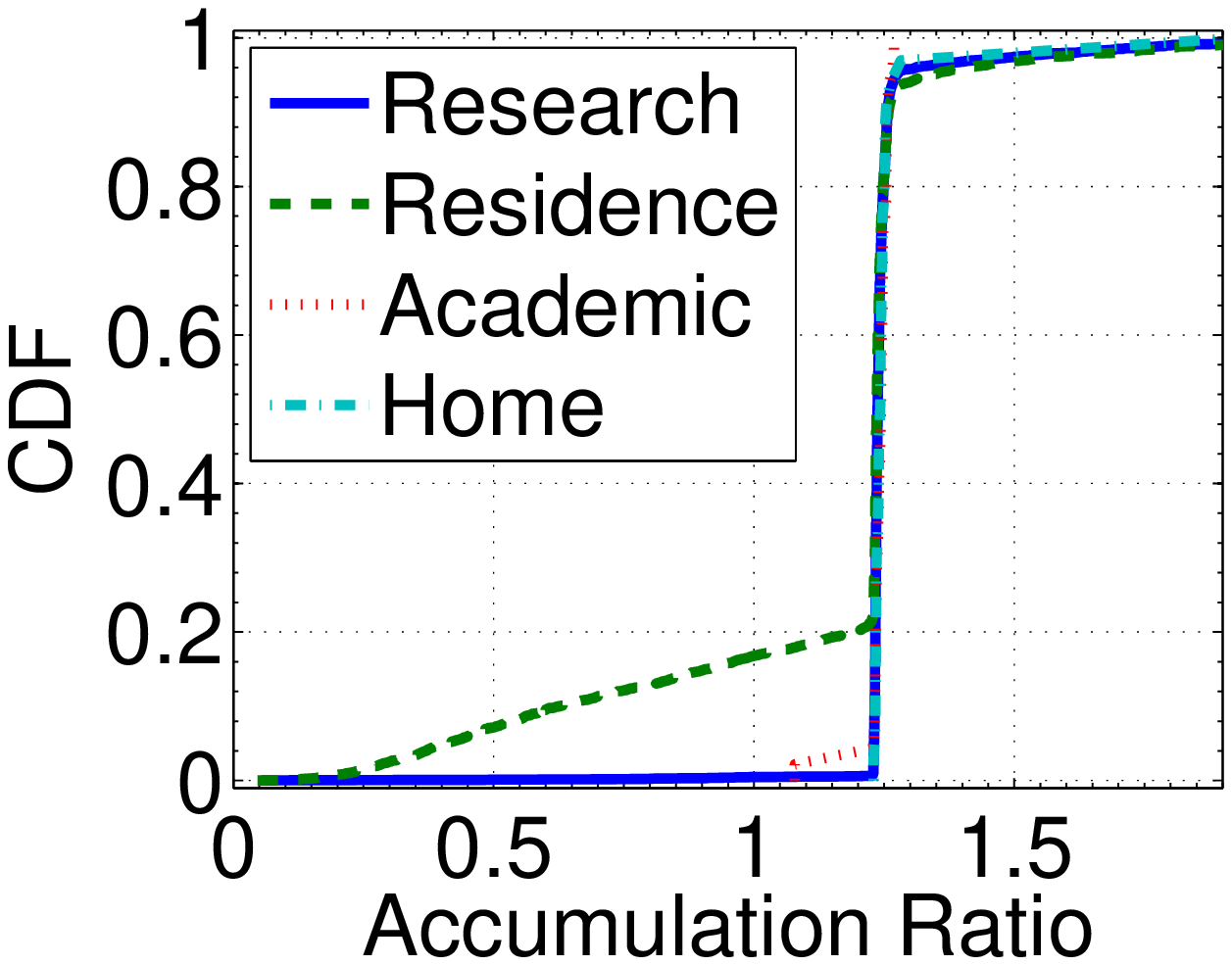}}
\caption{Steady State for Flash Videos. \emph{The server periodically
    transfers 64~kB of data to attain an accumulation ratio of 1.25
    (average download rate in steady state phase is 1.25 times the
    video encoding rate).}} 
\label{fig:SteadyShortONOFFFlash}
\end{figure}

\noindent
\textbf{ii) Steady State Phase.} We now show that YouTube servers
periodically transfer 64 kB blocks 
during the steady state phase to attain an accumulation ratio of 1.25
for Flash videos. In \autoref{fig:BlockSizeShortONOFFFlash} we present
the distribution of the block sizes observed while streaming videos in
the \emph{YouFlash} dataset; we observe that 64~kB is the dominant
block size in each network. The smaller block sizes observed in the
Residence and Academic networks are because of packet losses that
cause TCP retransmission timeouts. We observe block sizes larger than
64~kB when retransmissions due to packet losses merge multiple short
ON-OFF cycles to form a larger ON-OFF cycle. In
\autoref{fig:AccumulationRatioShortONOFFFlash} we observe an
accumulation ratio of approximately 1.25 for the majority of the
streaming sessions in each network.

\begin{figure}
  \centering
   \subfloat[Block Size.]{\label{fig:BlockSizeShortONOFFIEHtml5}\includegraphics[width=0.48\columnwidth]{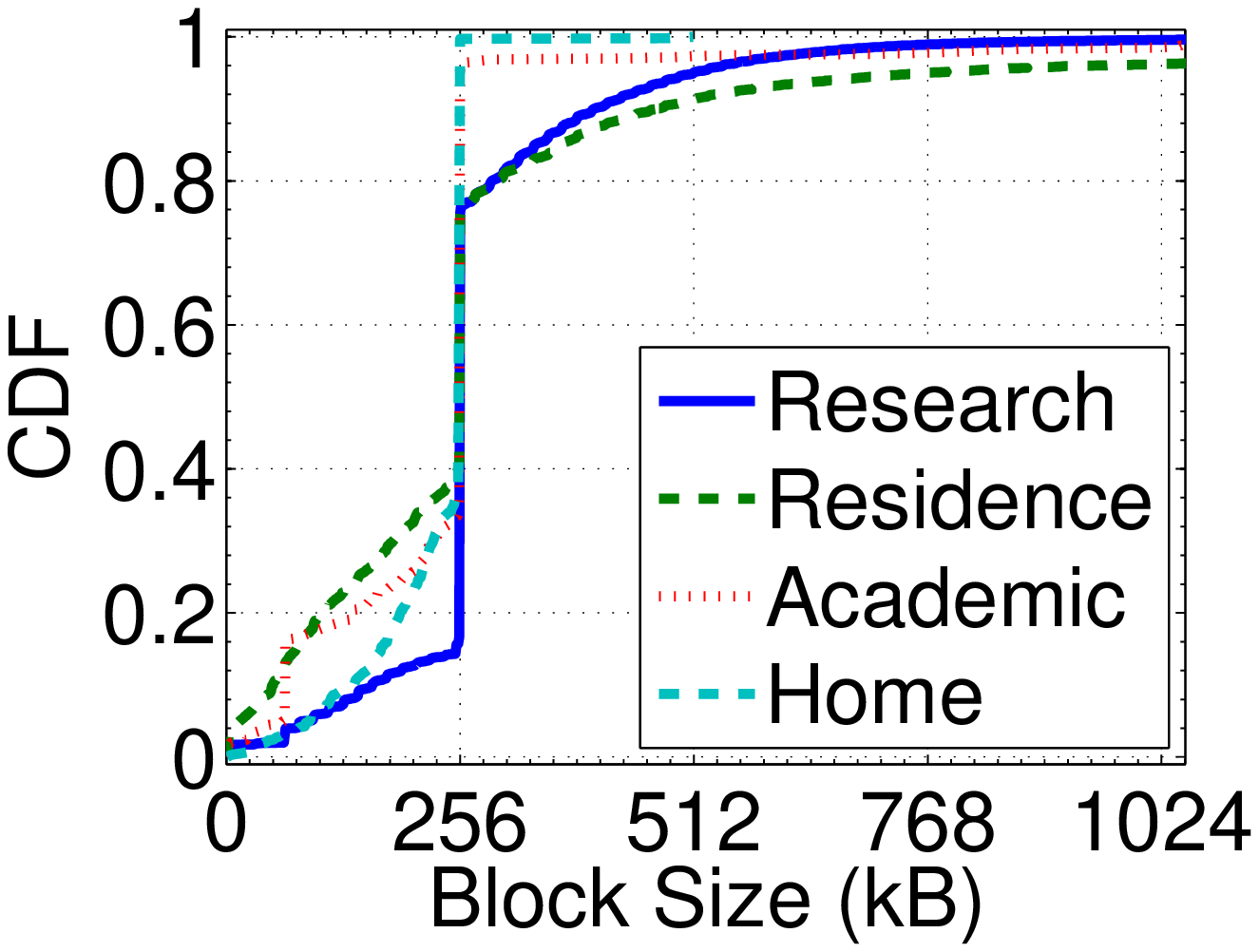}}
  \hspace{0.01\columnwidth}
  \subfloat[Accumulation Ratio. ]{\label{fig:AccumulationRatioShortONOFFIEHtml5}\includegraphics[width=0.48\columnwidth]{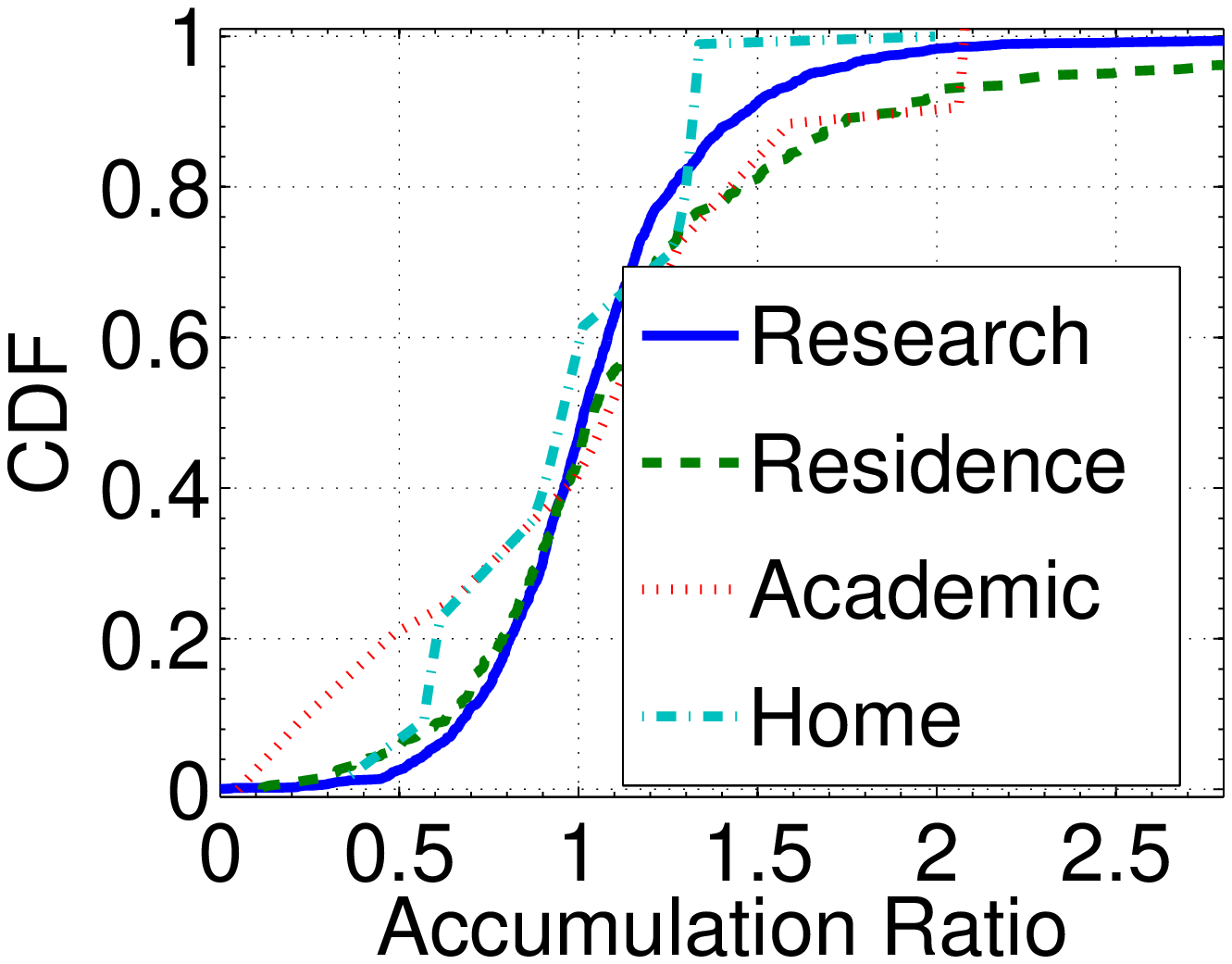}}
\caption{Steady State for HTML5 videos on Internet
  Explorer. \emph{A significant number of blocks have a size of
    256~kB. A wide range of accumulation ratios is observed while
    streaming HTML5 videos using Internet Explorer.}}
\label{fig:SteadyShortONOFFIEHtml5}
\end{figure}

For HTML5 on Internet Explorer, in
\autoref{fig:BlockSizeShortONOFFIEHtml5} we observe that 256 kB is the
dominant block size in each network. As in the case of Flash videos,
packet losses cause the block sizes to increase or decrease when
Internet Explorer is used to stream HTML5 videos. In
\autoref{fig:AccumulationRatioShortONOFFIEHtml5} we present the
distribution of the accumulation ratio when Internet Explorer is used
to stream HTML5 videos. In this figure, we observe a wide range of
accumulation ratios. We believe this wide range is an artifact of our
technique, or the technique used by the media player, to determine the
encoding rate of HTML5 videos. The mean and median accumulation ratio
across all the measurements presented in
\autoref{fig:AccumulationRatioShortONOFFIEHtml5} is 1.06 and 1.04
respectively.  
 
In summary, Flash videos, and HTML5 videos on Internet Explorer, use
short ON-OFF cycles. The dominant block size for Flash videos is 64~kB
and it is 256~kB for HTML5 on Internet Explorer.

\subsubsection{Long ON-OFF cycles}
\label{sec:YouTubeLong}

\begin{figure}
   \centering
    \subfloat[Download Amount and TCP Receive
      Window.]{\label{fig:DownLongONOFFGCHtml5}\includegraphics[width=0.48\columnwidth]{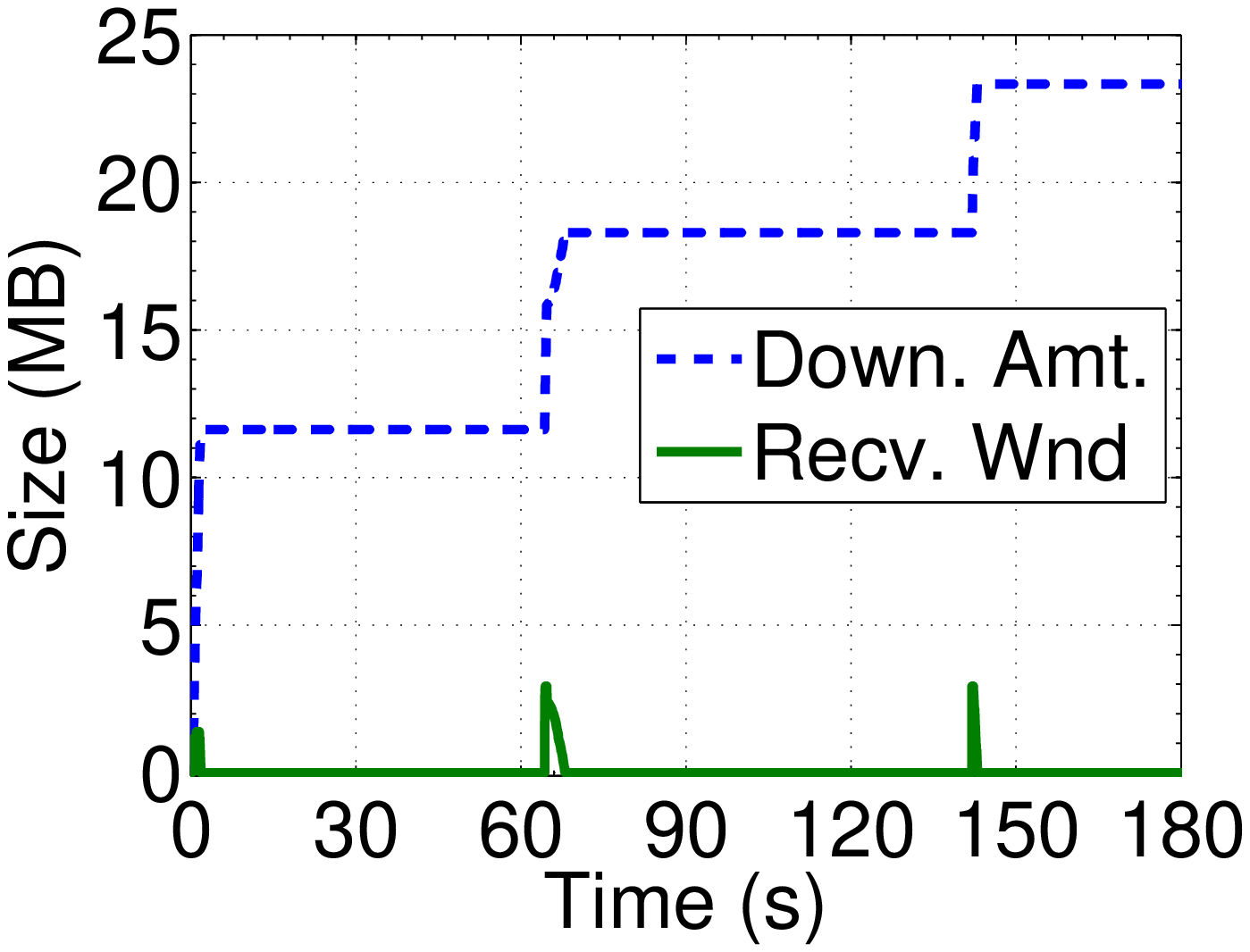}}
      \hspace{0.01\columnwidth}
      \subfloat[Block Size in the steady state phase.]{\label{fig:BlockLongONOFFHtml5}\includegraphics[width=0.48\columnwidth]{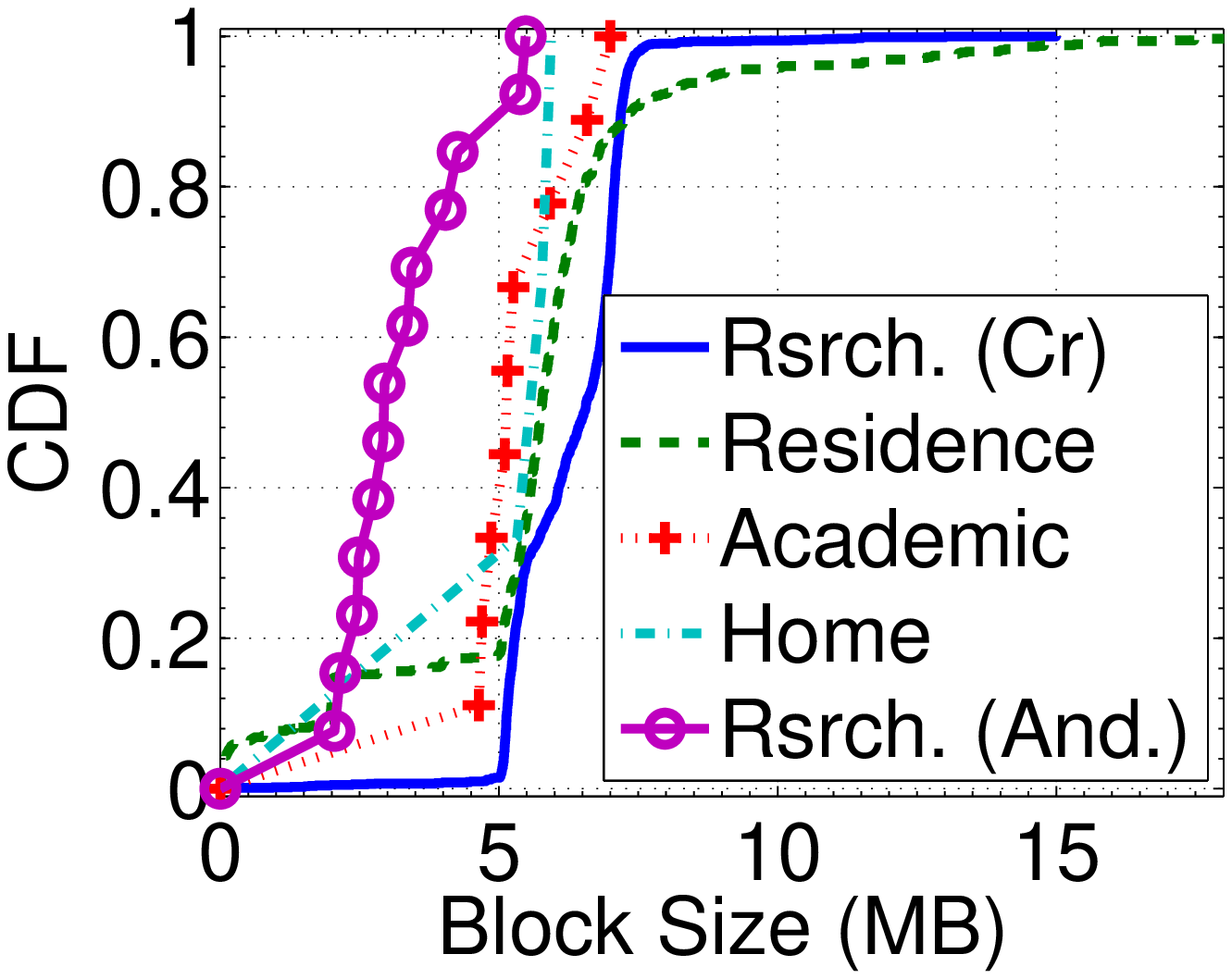}}
\caption{Long ON-OFF cycles. \emph{Long ON-OFF cycles are produced
    when blocks of large sizes are transferred in the steady state
    phase. Chrome browser periodically pulls large blocks
      resulting in long ON-OFF cycles.}}  
\label{fig:LongONOFF}
\end{figure}

In \autoref{fig:DownLongONOFFGCHtml5} we present a representative
trace for the long ON-OFF cycles. We observe OFF periods in the order
of 60 seconds during this measurement which was carried out in the
Research network using the Chrome browser. We observe that the TCP
receive window periodically becomes empty in
\autoref{fig:DownLongONOFFGCHtml5}. This shows that Chrome throttles
the data transfer rate by periodically pulling large blocks of data
resulting in long ON-OFF cycles. We make similar observations when
HTML5 videos are streamed using the native YouTube application for Android
devices. 

We now present our observations on the buffering phase and the steady
state phase when long ON-OFF cycles are used to stream YouTube videos;
we used the videos in the \emph{YouHtml} and \emph{YouMob} dataset for
our measurements.

\noindent \textbf{i) Buffering Phase.} During our measurements we
observed that the amount downloaded during the buffering phase by Chrome
and Android is independent of the video encoding rate. In each
network, we observe a scatter plot similar to the one presented for
Internet Explorer in \autoref{fig:BufferingShortONOFFIEHtml5}. We do
not present these figures due to lack of space. We observe that while
Chrome typically downloads between 10~MB and 15~MB during the
buffering phase, the native YouTube application for Android downloads
from 4~MB to 8~MB during the buffering phase.

\noindent
\textbf{ii) Steady State Phase.} In \autoref{fig:BlockLongONOFFHtml5}
we present the distribution of the block sizes when long ON-OFF cycles
were observed when streaming YouTube videos. In this figure we
observe block sizes larger than 2.5~MB for most of the streaming
sessions; the measurements carried out in the Research network using
Chrome and Android are presented as \emph{Rsrch. (Cr)}, and
\emph{Rsrch. (And.)}, respectively in
\autoref{fig:BlockLongONOFFHtml5}. 
 
We observe a mean and median accumulation ratio of 1.34 and 1.29 for
Chrome, and 1.24 and 1.15 for Android in the Research networks. We
do not present these results because we observe a wide range of
accumulation ratios. We believe this wide range to be an artifact of
our estimation process of the video encoding rate. 

In summary, Chrome and Android periodically pull blocks that have a
size larger than 2.5~MB to throttle the download rate in the steady
state phase.

\subsubsection{Combination of ON-OFF Strategies}
\label{sec:YouTubeCombONOFF}

We now use two videos from the \emph{YouMob} dataset to show that iPad
uses more than one strategy for streaming YouTube videos; we call
these two videos Video1 and Video2.

\begin{figure}
  \centering
   \subfloat[Download Evolution.]{\label{fig:DownCombONOFF1}\includegraphics[width=0.48\columnwidth]{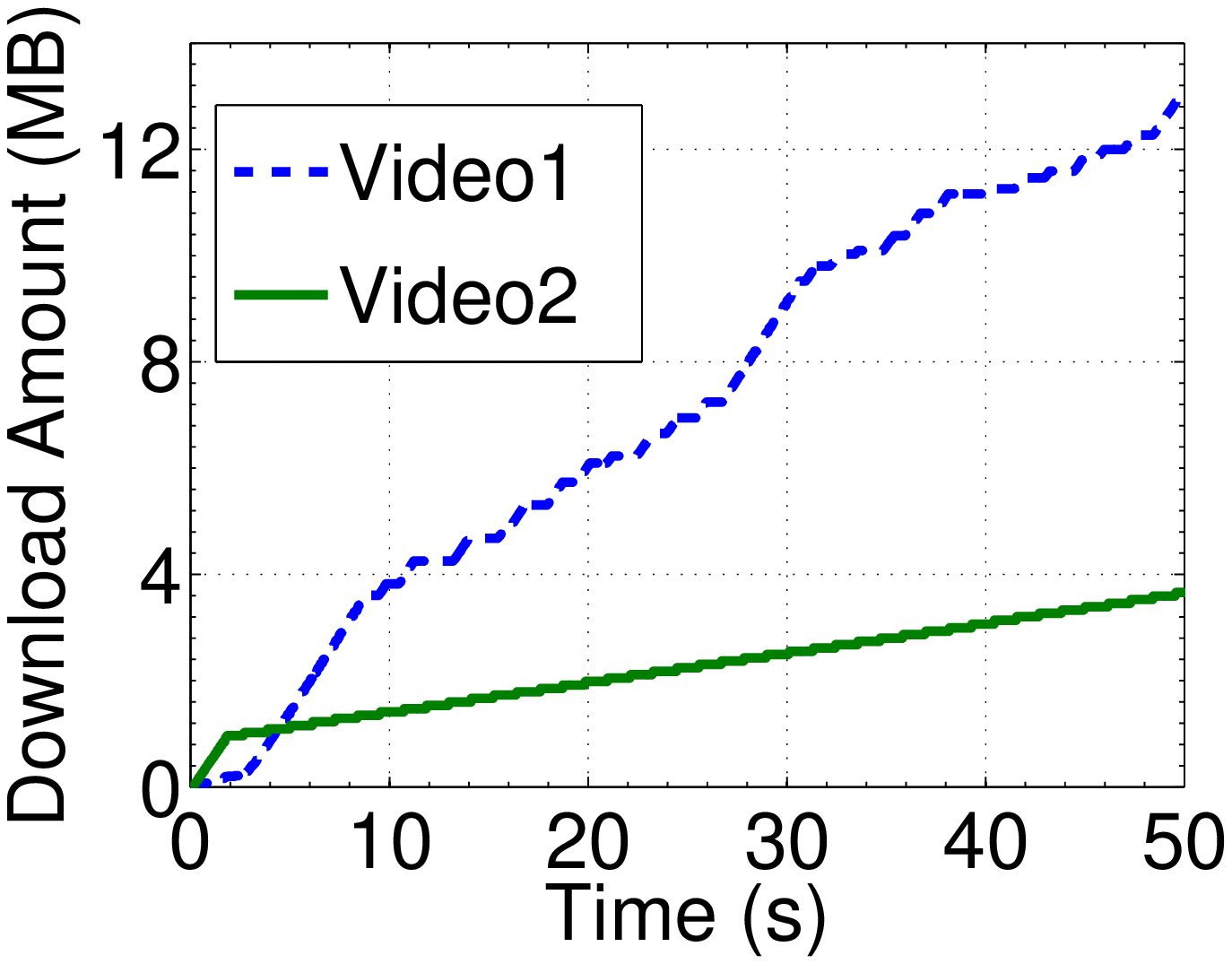}}
  \hspace{0.01\columnwidth}
  \subfloat[Block Size and Encoding Rate.]{\label{fig:CombONOFFBlockReason}\includegraphics[width=0.48\columnwidth]{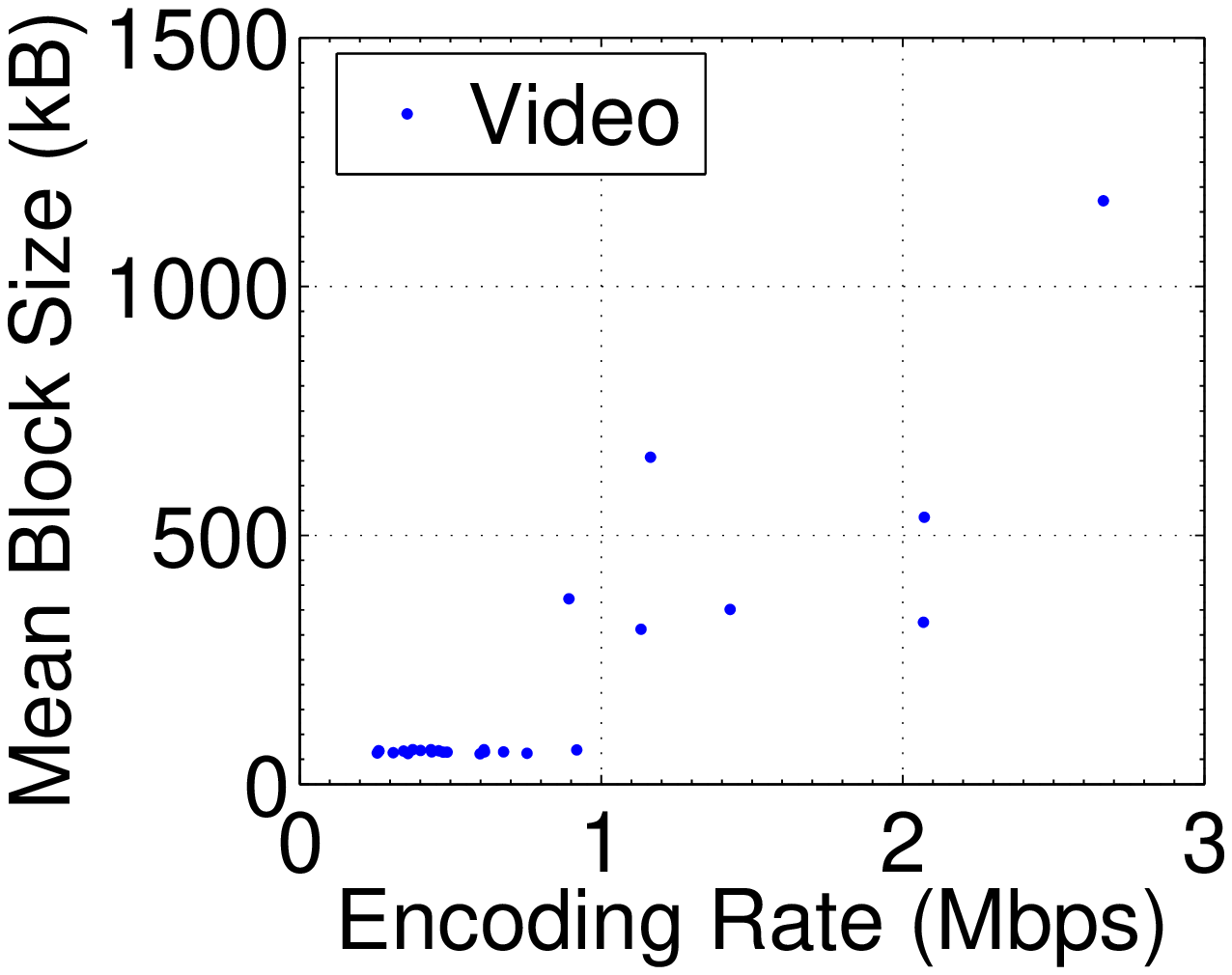}}\\
  \caption{Different streaming strategies for YouTube videos on iPad.} 
  \label{fig:CombONOFF}
\end{figure}

For Video1, in \autoref{fig:DownCombONOFF1}, we observe periodic
buffering followed by short ON-OFF cycles. Further, we observe that 37
different TCP connections were successively used for the data transfer
in the first 60 seconds. For each connection, the HTTP {\tt GET}
request contained the range of data requested for a given
connection. The amount of data transferred in each TCP connection
varied from 64~kB to 8~MB. In comparison, for Video2, we observe short
ON-OFF cycles in \autoref{fig:DownCombONOFF1}; only one TCP
connection was used to transfer the video contents.  

In \autoref{fig:CombONOFFBlockReason} we observe that the block
size used during a streaming session depends on the video encoding
rate. A given YouTube video may be available in multiple resolutions
and the native YouTube application chooses a resolution according to
available network and device
capabilities~\cite{Finamore_2011_YouTubeDeviceInfra}. This implies
that streaming strategies for mobile devices with large screens such
as the iPad may depend on end-to-end available bandwidth. The measurements
presented in \autoref{fig:CombONOFFBlockReason} were carried out in
the Research network which has sufficient bandwidth for streaming high
resolution videos. 

In summary, we observe that the streaming strategy depends on the
encoding rate, and the end-to-end available bandwidth, for an
iPad. 

\subsubsection{No ON-OFF Cycles}
\label{sec:YouTubeNOONOFF}

\begin{figure}
  \begin{minipage}[t]{0.48\columnwidth}
  \centering
  \includegraphics[width=\columnwidth]{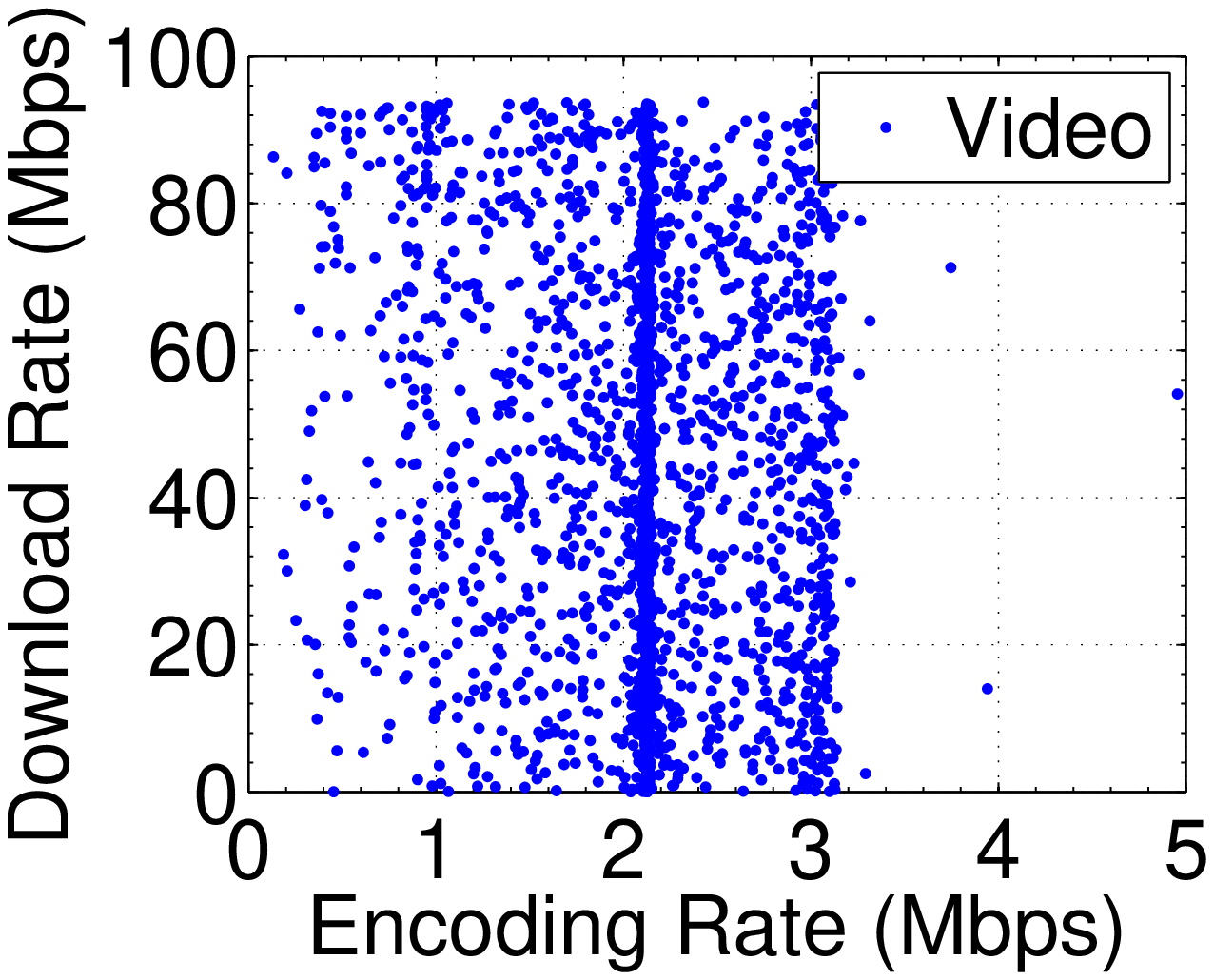}
  \caption{No ON-OFF Cycles.}
  \label{fig:NoOnOFFCorr} 
  \hspace{0.01\columnwidth}
  \end{minipage}
  \begin{minipage}[t]{0.48\columnwidth}
  \centering
  \includegraphics[width=\columnwidth]{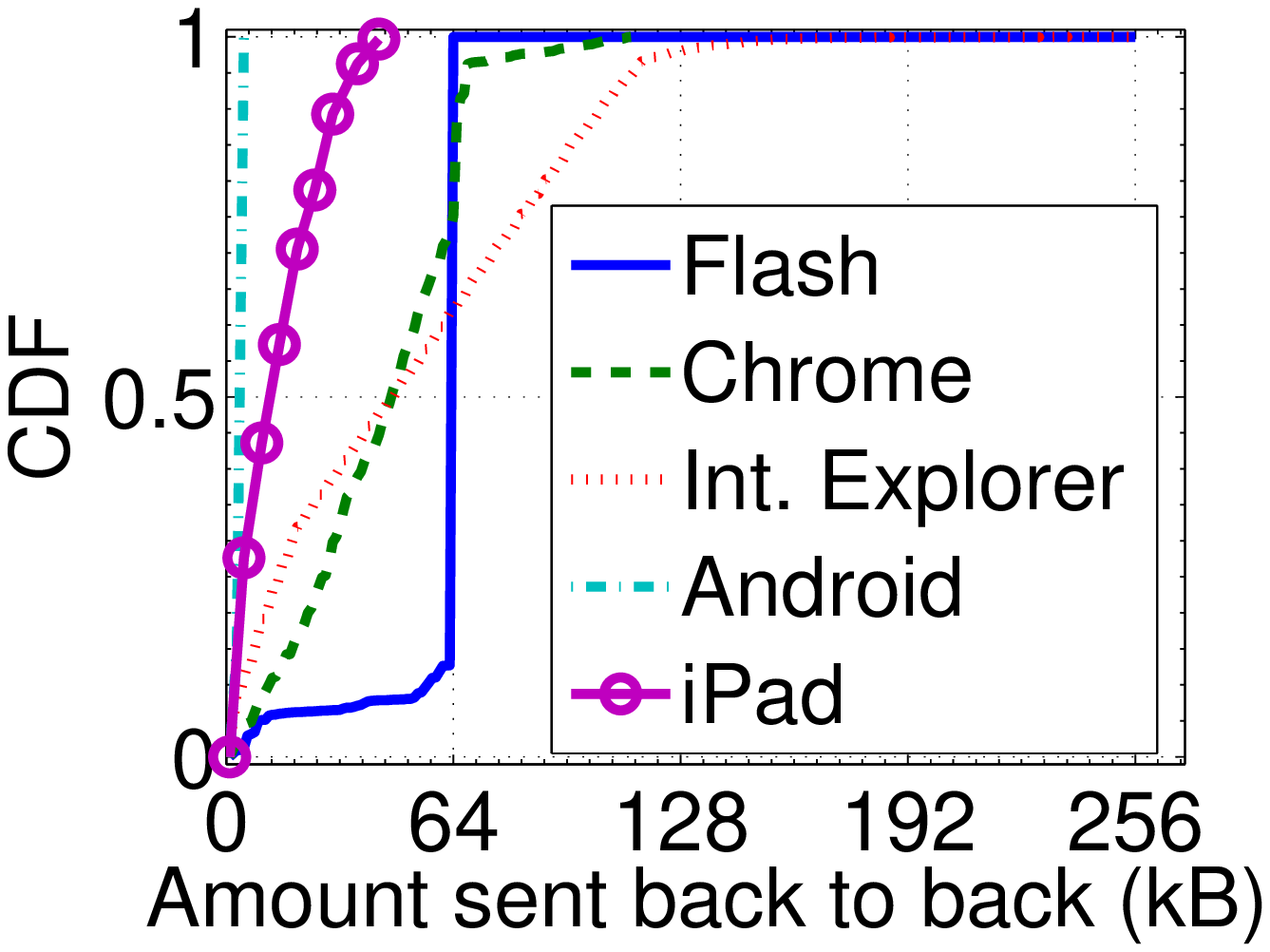}
  \caption{Ack Clock.}
   \label{fig:AckClockShortONOFFFlash}
  \end{minipage}
\end{figure}

We observe the streaming strategy of no ON-OFF cycles when neither the
server nor the client limit the rate of data transfer. The whole
video is downloaded during the buffering phase; such video streaming
sessions do not contain a steady state phase. This streaming strategy
is observed for the following two cases: HTML5 videos on Firefox, and
for Flash HD videos.  

In \autoref{fig:NoOnOFFCorr} we observe that the
download rate for HD videos is not correlated to the encoding rate. We
make a similar observation for HTML5 videos on Firefox. The
measurements presented in \autoref{fig:NoOnOFFCorr} were 
carried out in the Research network. We made similar observations for
the other networks used in our measurements. To ensure that this
behavior is not due to a large buffering phase, we selected 50 HD
videos and 50 HTML5 videos from the \emph{YouHD} and \emph{YouHtml}
dataset that have a duration larger than 1200 seconds. For each of
these videos, we did not observe a steady state phase during the entire
streaming session.   

\subsubsection{Discussion on ACK Clocks}

TCP is an ack-clocked
protocol~\cite{Jacobson_1988_CongAvoidControl}. The ACK clock enables
the TCP source to estimate the end-to-end available bandwidth before
sending a packet. This estimate is used to determine the size of the
TCP congestion window. Allman~\etal~\cite{RFC5681_2009_TCPCC} suggest
that the TCP congestion window be reset after idle periods in the
order of a retransmission timeout. This reset ensures that the TCP
source does not overwhelm the network without probing the end-to-end
available bandwidth.

In \autoref{fig:AckClockShortONOFFFlash} we present the distribution
of the amount of data received during the first round-trip time 
of the ON periods in the Research Network. This amount is a
conservative estimate of the TCP congestion window at the beginning of
an ON period. For short ON-OFF cycles we observed OFF periods of
duration between 0.2~seconds to 5~seconds while for the long ON-OFF
cycles we observed OFF periods up to 80~seconds long. In
\autoref{fig:AckClockShortONOFFFlash} we observe that the congestion
window is not reset after the OFF periods. For example, for Flash
videos, we observe that the entire block of 64~kB is sent without
probing the end-to-end available bandwidth. The curves in
\autoref{fig:AckClockShortONOFFFlash} represent the minimum of the TCP
congestion window and the block size used during the steady state
phase. Because the block size depends on the streaming strategy, and
thus the application, we observe different curves for each application
in \autoref{fig:AckClockShortONOFFFlash}.

This observation is important as the absence of an ack-clock can increase
the loss rate in the networks.

\subsection{Netflix Streaming Strategies}
\label{sec:Netflix}

\begin{figure}
  \centering
   \subfloat[Short ON-OFF.]{\label{fig:NetflixShortONOFF}\includegraphics[width=0.48\columnwidth]{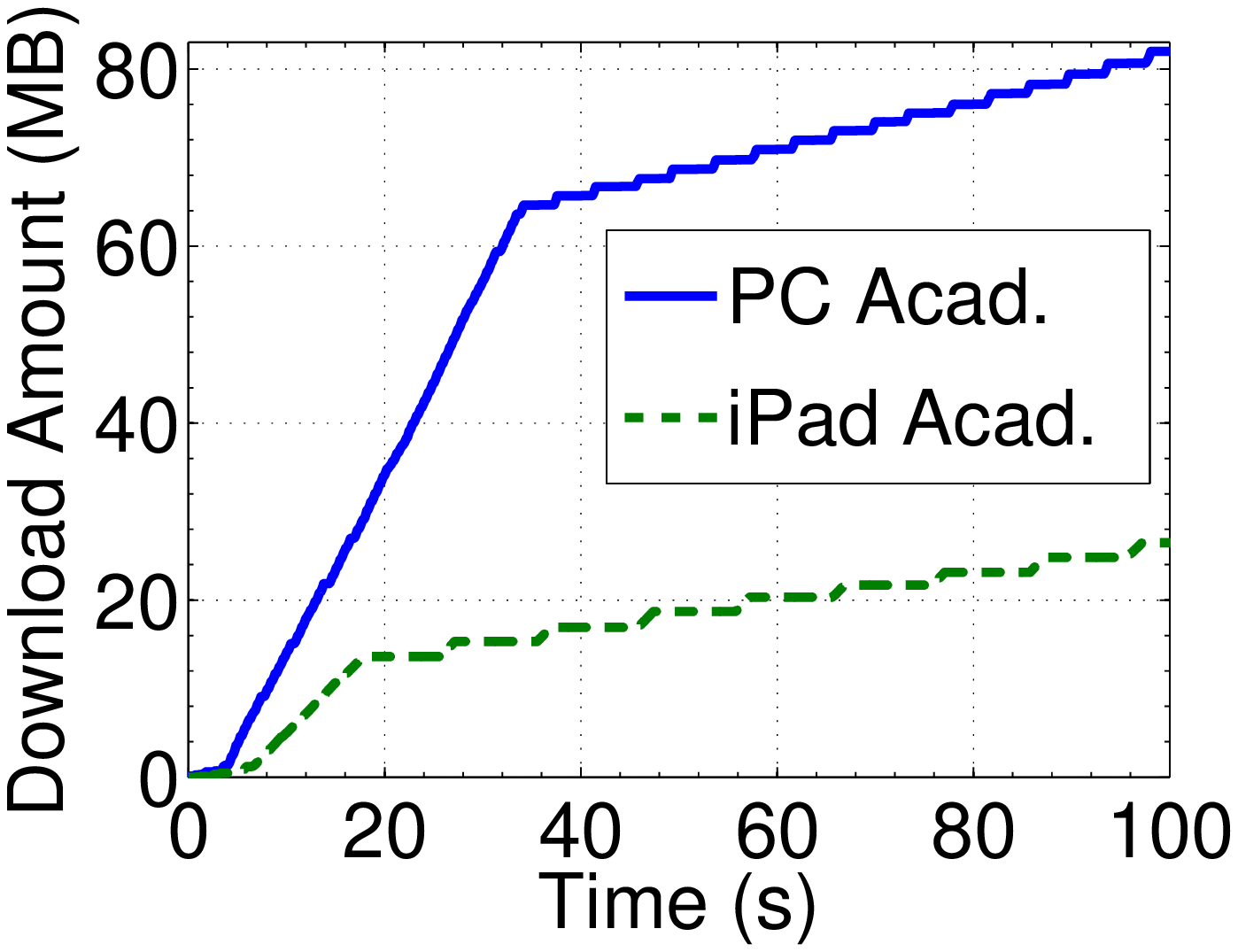}}
  \hspace{0.01\columnwidth}
  \subfloat[Long ON-OFF.]{\label{fig:NetflixLongONOFF}\includegraphics[width=0.48\columnwidth]{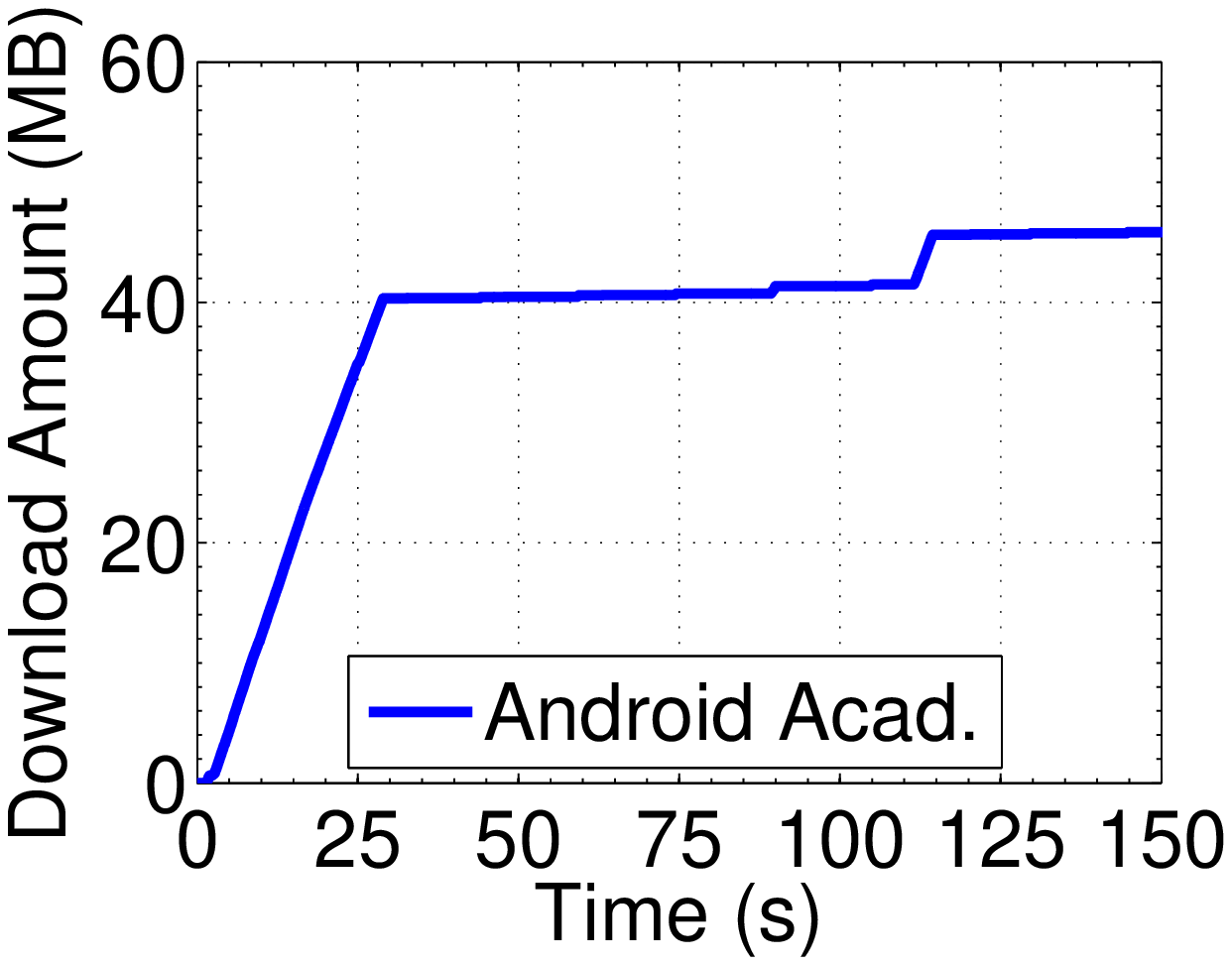}}\\
  \caption{Streaming Strategies used by Netflix. \emph{Short ON-OFF
      cycles for PCs and iPad. Long ON-OFF cycles used for the Android 
      application.}} 
  \label{fig:NetflixStrategy}
\end{figure}

We now use one video from the \emph{NetPC} dataset and one video from
the \emph{NetMob} dataset to provide an overview of the strategies
used to stream Netflix videos. \autoref{fig:NetflixShortONOFF}
presents the evolution of the download amount for the first 100
seconds from the beginning of video download. In
\autoref{fig:NetflixShortONOFF} we observe short ON-OFF cycles when
Web browsers and the native application for the iPad is used to stream
the Netflix videos. In \autoref{fig:NetflixLongONOFF}, we observe long
ON-OFF cycles when Netflix videos are viewed using the native mobile
application for Android. The measurements presented in
\autoref{fig:NetflixStrategy} were carried out in the Academic
Network. 

We now use the videos in \emph{NetPC} and \emph{NetMob} dataset to
study the buffering and steady state phases.

\subsubsection{Buffering Phase}

\begin{figure}
  \centering
   \subfloat[Short ON-OFF.]{\label{fig:BufferingNetflixShortONOFF}\includegraphics[width=0.48\columnwidth]{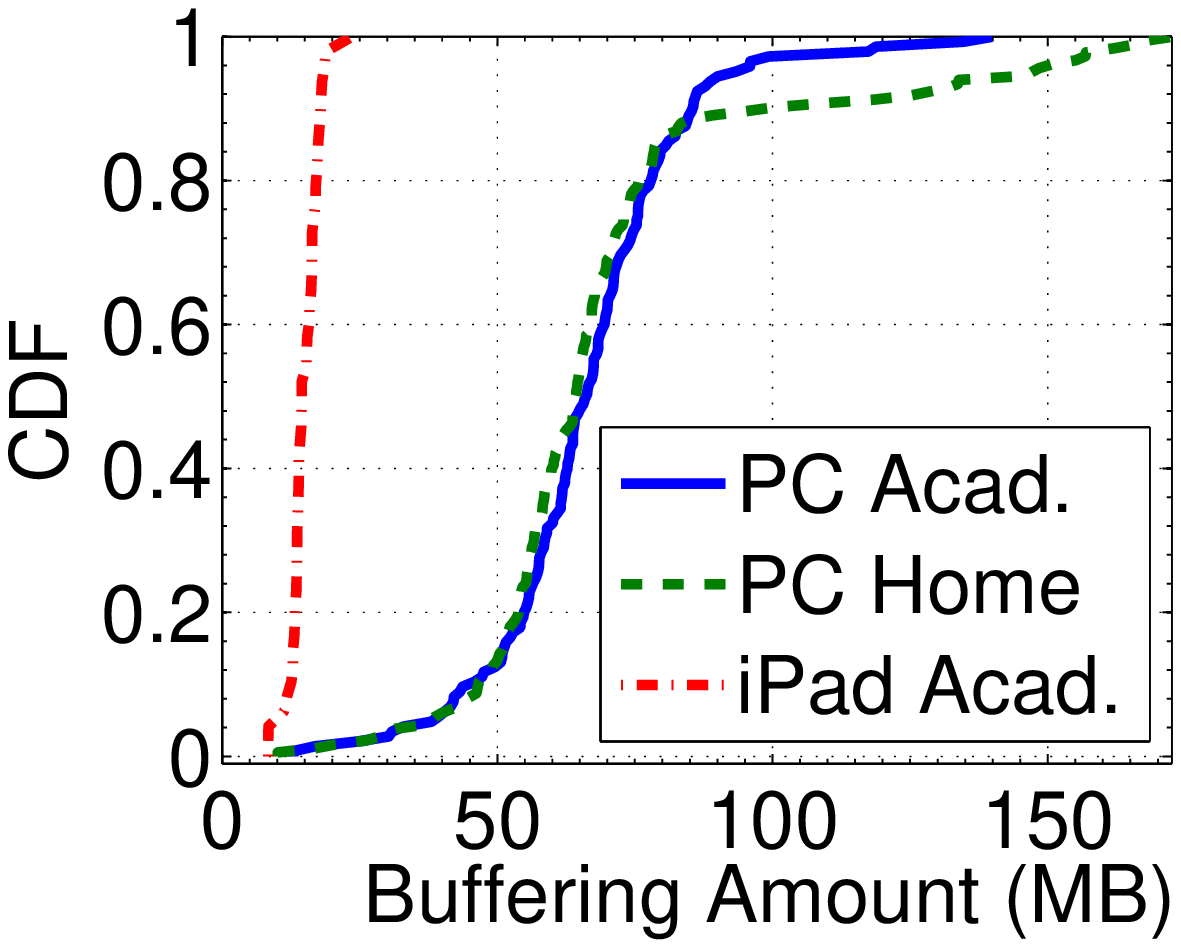}}
  \hspace{0.01\columnwidth}
  \subfloat[Long ON-OFF.]{\label{fig:BufferingNetflixLongONOFF}\includegraphics[width=0.48\columnwidth]{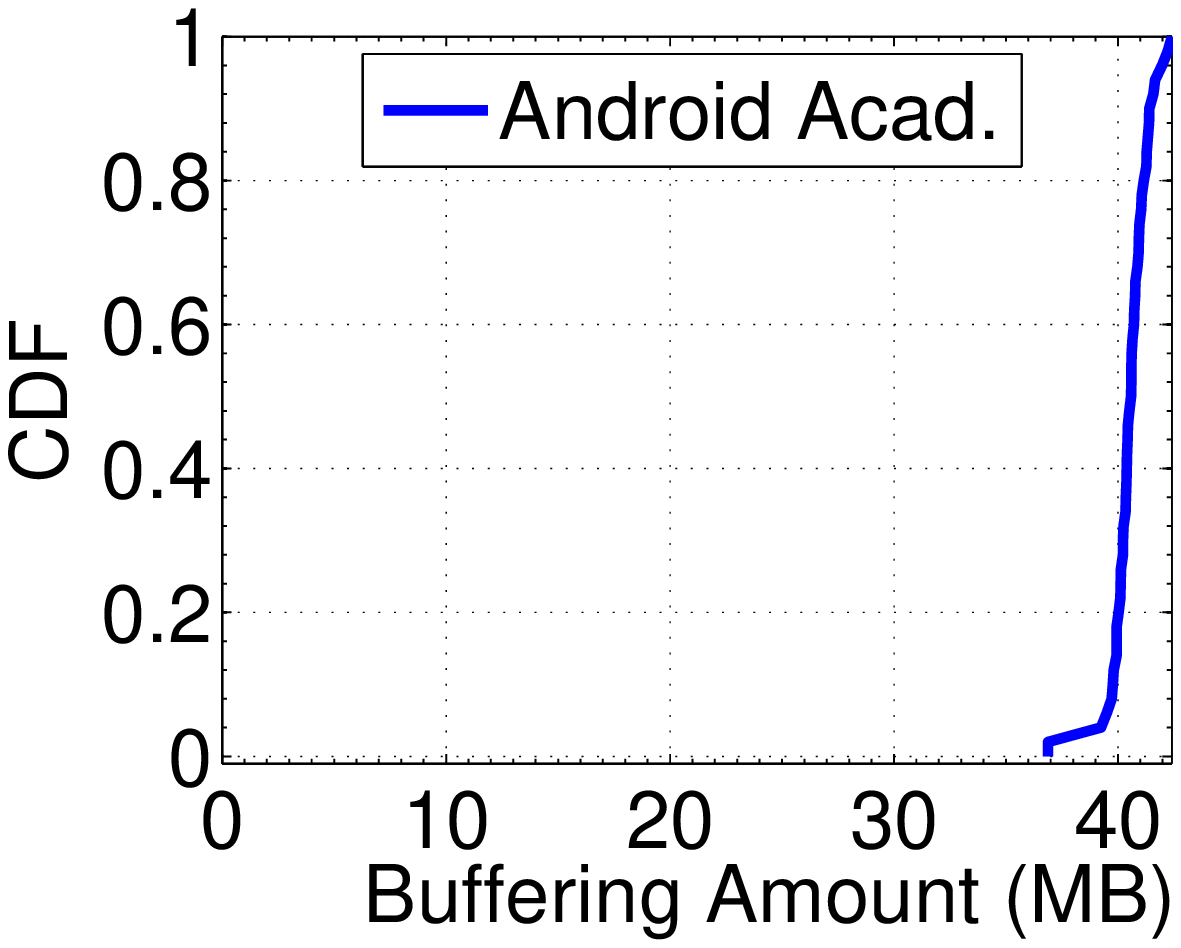}}\\
  \caption{Buffering Amount. \emph{Netflix transfers multiple copies of
        the same video content at different encoding rates during the
        buffering phase}.}  
  \label{fig:NetflixBuffering}
\end{figure}

In \autoref{fig:NetflixBuffering}, we observe that the amount
downloaded during the buffering phase depends on the application, Web
browser (for PCs) or the native mobile application. In
\autoref{fig:BufferingNetflixShortONOFF}, for PCs we observe download
amounts in the order of 50~MB; however, for the native iPad application we
observe download amounts in the order of 10~MB. We now present a
possible reason for this behavior. Each Netflix video is available in
different encoding
rates. Akhshabi~\etal~\cite{Akhshabi_2011_StreamingRateAdaption} show  
that when a Netflix streaming session begins, the video fragments of
all the available encoding rates are downloaded during the buffering
phase. We hypothesize that the encoding rates for an iPad may be
selected from a subset of available encoding rates. In
\autoref{fig:BufferingNetflixLongONOFF} we observe that the amount of
data downloaded during the buffering phase by the native Android application is
in the order of 40~MB. This is significantly larger than what we
observe for the native iPad application.

\subsubsection{Steady State Phase}

\begin{figure}
  \centering
   \subfloat[Short ON-OFF.]{\label{fig:BlockSizeNetflixShortONOFF}\includegraphics[width=0.48\columnwidth]{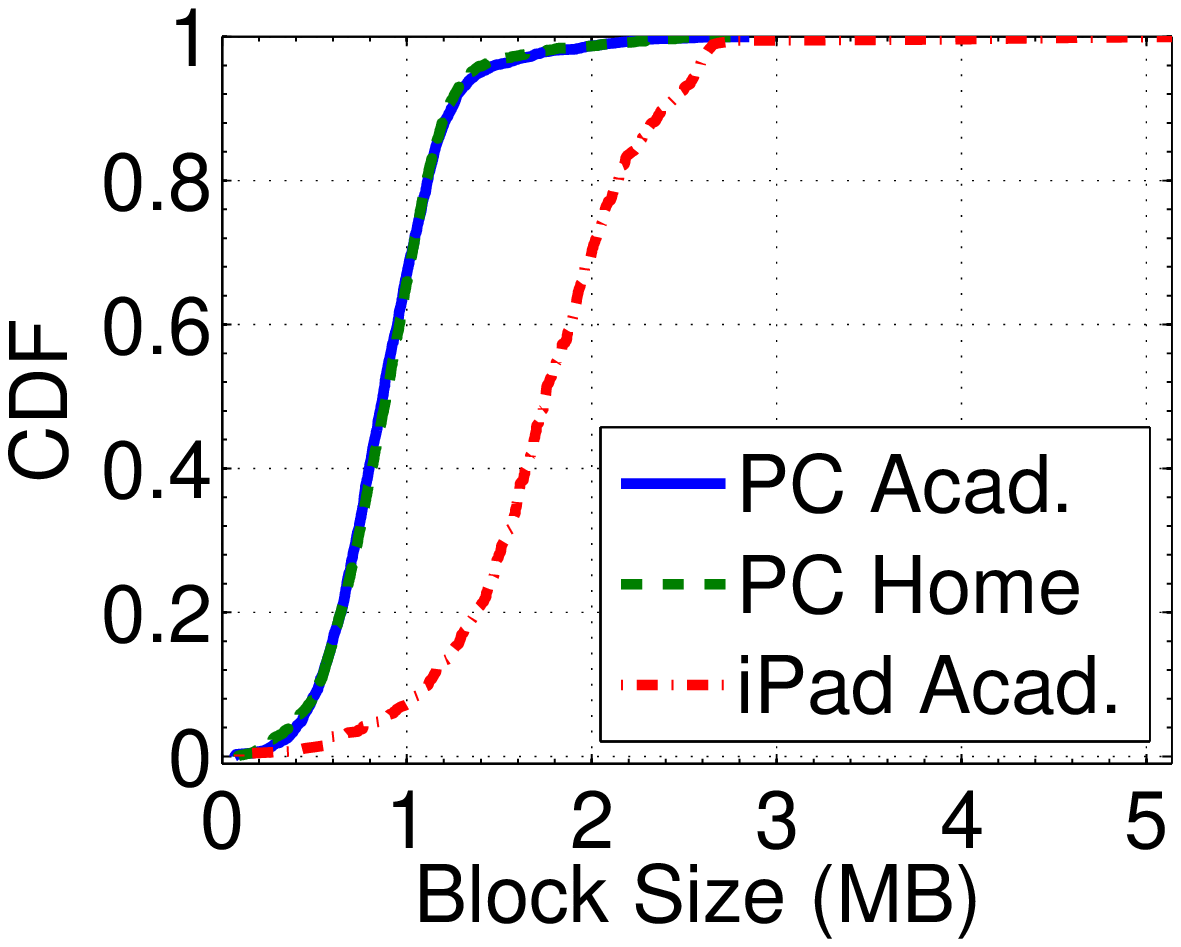}}
  \hspace{0.01\columnwidth}
  \subfloat[Long ON-OFF.]{\label{fig:BlockSizeNetflixLongONOFF}\includegraphics[width=0.48\columnwidth]{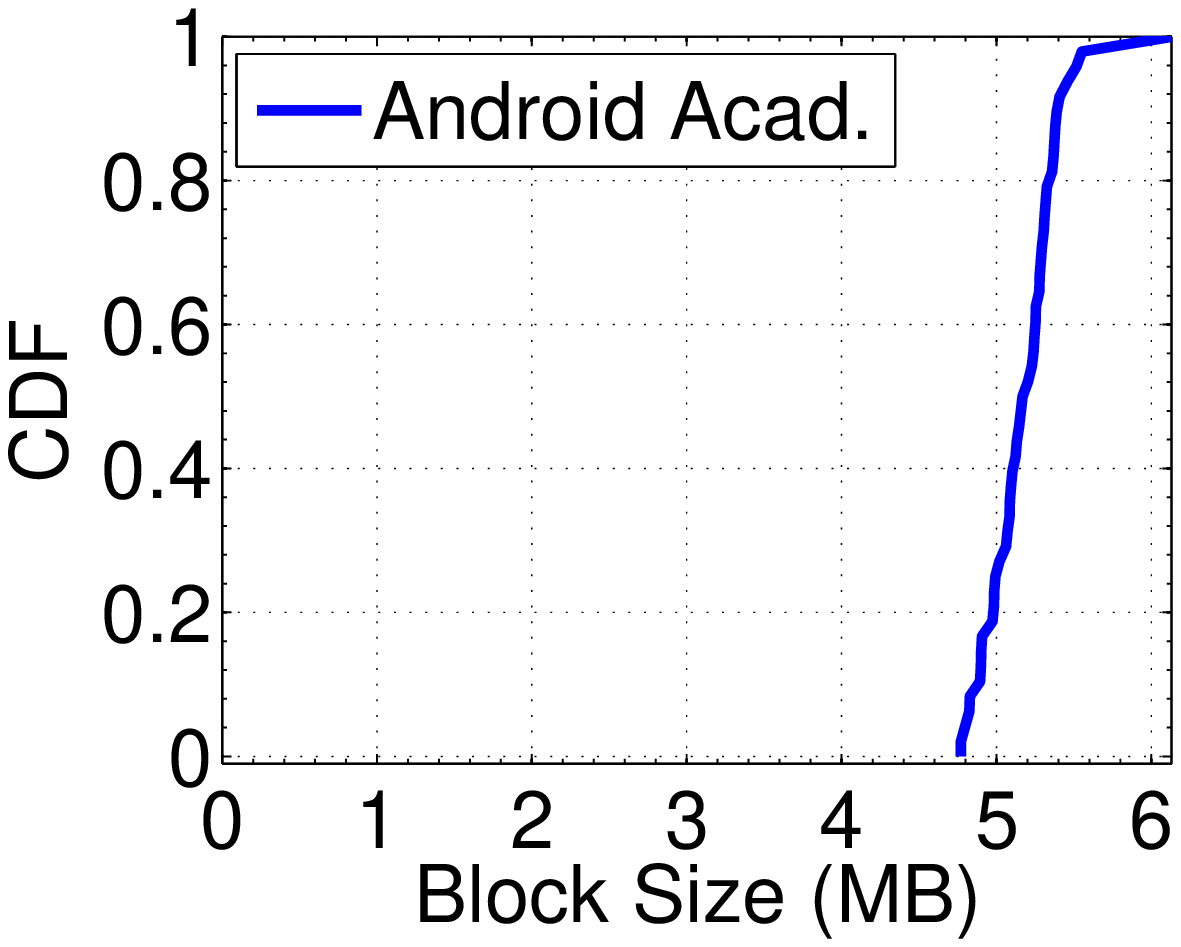}}\\
  \caption{Distribution of block sizes for Netflix
    videos. \emph{The block sizes during Netflix streaming
        sessions depend on the application used.}} 
  \label{fig:NetflixBlockSize}
\end{figure}

In \autoref{fig:BlockSizeNetflixShortONOFF} and
\autoref{fig:BlockSizeNetflixLongONOFF} we observe that the block
sizes used to stream Netflix videos depend on the application, Web
browser or the native mobile application. For example, we observe
large blocks when Netflix videos are streamed using the native Android
application. These large blocks produce long ON-OFF cycles such as
those observed in \autoref{fig:NetflixLongONOFF}. For the strategy of
short ON-OFF cycles, we observe that the majority of the blocks have a
size smaller than 2.5~MB. These blocks are however slightly larger
than the 64~kB and 256~kB blocks we observed when short ON-OFF cycles
were used to stream YouTube videos.  

During our measurements we observed that Netflix uses a large number
of TCP connections to transfer the video contents to PCs and the
iPad. We are currently not able to speculate the reasons for this
behavior. We observed ack-clocks when a new connection was used to
send a block of data; we did not observe ack-clocks when a connection
was used to send more than one block. We observe ack clocks when
Netflix videos are streamed using the native mobile application for
Android. We have not presented these results due to lack of space. 

\subsection{Discussion}
\label{sec:SummaryYouTubeNetflix}

\begin{table}
\begin{footnotesize}
\begin{tabular}{|>{\centering} m{0.25\columnwidth} |
    >{\centering}m{0.18\columnwidth} |
    >{\centering}m{0.18\columnwidth} |
    >{\centering}m{0.18\columnwidth} |}
\hline
Strategy & No ON-OFF & Long ON-OFF & Short ON-OFF
\tabularnewline
\hline
Engineering Complexity & Not required &
\multicolumn{2}{m{0.42\columnwidth}|}{Explicit support at Application Layer} 
\tabularnewline
\hline
Receive buffer occupancy & Large & Moderate & Small 
\tabularnewline
\hline
Unused bytes on user interruption & Large amount  & Moderate amount & Small amount
\tabularnewline
\hline
\end{tabular}
\end{footnotesize}
\caption{Comparison of streaming strategies.}
\label{tab:CompStreaming}
\end{table}

In this section we detail the network characteristics of YouTube and
Netflix traffic and show that strategies used to stream YouTube and
Netflix videos depend on the application and the container. For
the Flash container, we observe the streaming strategy is independent
of the application used. This is because the applications do not
control the data transfer rate; rate control, if any, is performed by
the YouTube servers. However the YouTube servers do not limit the   
data transfer rate when streaming HTML5 videos; each application uses
its own strategy to stream HTML5 videos. We therefore 
observe a wide range of patterns for the buffering phase and the
steady state phase for HTML5 videos. When Netflix videos are streamed
to Web browsers we observe the same streaming strategy regardless
of the of the Web browser. However, the strategy is different for the
native mobile application for the Android device and the iOS
device. 

In \autoref{tab:CompStreaming} we summarize the difference
between the three streaming strategies. 

Of the three streaming strategies we identified, the streaming
strategies of no ON-OFF cycles is a TCP file transfer. Therefore, we
believe that this strategy requires no complex engineering at the
application layer. The other two streaming strategies, short ON-OFF
cycles and long ON-OFF cycles, explicitly restrict the data transfer
rate at the application layer. We therefore believe that engineering
is required at the application layer for the strategies of short
ON-OFF cycles and long ON-OFF cycles. 

The strategies of short ON-OFF cycles and long ON-OFF cycles achieve
their goals by restricting the amount downloaded during the buffering phase
followed by restricting the data transfer rate according to a desired
accumulation ratio. A small accumulation ratio and buffering amount is
desirable because it reduces the amount of unused bytes in the buffers
of the players. This is also important for mobile devices that may
have storage constraints. The amount of unused bytes is also important
because recent studies have shown that users tend to interrupt the
video download due to lack of interest
\cite{Finamore_2011_YouTubeDeviceInfra, Gill_2007_Youtube,
  Huang_2007_VODProfit}.  

The strategies of short ON-OFF cycles and long ON-OFF cycles limit the
data transfer rate at the application layer. We show that the traffic
characteristics while using these strategies might not be the one of a
standard TCP flow. For example, we show the absence of ack-clocks in
the TCP connections used to stream Netflix and YouTube videos.

We can therefore conclude that migration from one application to
another, or from one container to another, can impact the aggregate
video streaming traffic. For example, migration from Flash to HTML5, and
increase in the usage of mobile devices are two possibilities
that cannot be ruled out. We present a mathematical model to study
this impact in the following section.

\section{Model for Aggregate Video Traffic}
\label{sec:Model}
In \autoref{sec:YouTube} and \autoref{sec:Netflix}, we show that the
application and the container determine the strategy to stream
videos. In this section, we present a mathematical model to express
the stochastic properties of the aggregate video streaming traffic as
a function of the video parameters. Our model can be used to dimension
the network and quantify the impact of migrating from one strategy to
another. We first develop our model for the case of users that do not  
interrupt the video download. We then study the impact of user
interruption due to lack of interest on the accumulation ratio and the
amount of data downloaded in the buffering phase. We then quantify the
amount of bandwidth wasted when users interrupt the video download due to
lack of interest.
 
For our model, we assume that the video streaming sessions arrive
according to a homogeneous Poisson process with rate $\lambda$. We use
the measurements performed by Yu~\etal~\cite{Yu_2006_UserBehaviorVOD}
for the Poisson assumption of the arrival rate\footnote{Given the fact
  that users watch the videos in series, it is easy to prove that the
  Poisson assumption is not needed at the video level. It is enough to
  have the Poisson assumption at the user level, which is very likely
  to be the case given the human nature of this activity.}. Let $T_n$,
$n \in \mathds{Z}$, denote the arrival time of the $n$-th video. We
assume that $n-$th video is streamed at a fixed encoding rate, $e_n$,
and has a fixed duration (length), $L_n$; the size of the $n$-th video
is $S_n = e_nL_n$. We also assume that the network is over
provisioned: the end-to-end available bandwidth is larger than the
video encoding rate for each video streaming session. This hypothesis
is validated by our measurements presented in
\autoref{sec:YouTubeShort}. Indeed, for the videos in \emph{YouFlash}
dataset we observed an accumulation ratio larger than one, which
implies that the download rate, and hence the end-to-end available
bandwidth, is larger than the video encoding rate.
 
\subsection{Video Download without Interruptions}
\label{sec:ModelNoInterrupt}

We now model the aggregate data rate of video streaming traffic when
users do not interrupt the video download.  We first examine the
strategy of no ON-OFF cycles where the whole video is downloaded at
the end-to-end available bandwidth. We assume the time required to
download the $n$-th video is $D_n$. For the $n$-th video, the video
download is \emph{active} at time $t$ when $T_n \le t \le T_n +
D_n$. Let $X_n(t-T_n)$ denote the download rate of the $n$-th video at
time $t$; $X_n(t) = 0$ when $t < T_n$ and $t > T_n+D_n$. Let $R(t)$
denote the aggregate data rate of the video streaming traffic at time
$t$.

According to
Barakat~\etal~\cite{Barakat_2002_FlowModelInternetBackbone}, the mean
and variance of the aggregate data rate are:
\begin{align}
&\mathds{E}[R(t)] = \lambda \mathds{E}[S_n], \label{eqn:MeanPaper}\\
&V_R = \mathds{E}[R^{2}(t)] - (\mathds{E}[R(t)])^{2}= \lambda \mathds{E}[\int_{0}^{D_n}X_{n}^{2}(u)du],\label{eqn:VarPaper}
\end{align}
respectively 

When the download rate of the $n$-th video is a constant $G_n$,
substituting $D_n = \dfrac{S_n}{G_n}$, $S_n=e_nL_n$, and $X_n(t)=G_n$
for $T_n \le t \le T_n+D_n$, in equations~\ref{eqn:MeanPaper} and
\ref{eqn:VarPaper} yields:
\begin{eqnarray}
\mathds{E}[R(t)] &=& \lambda \mathds{E}[e_n]\mathds{E}[L_n], \label{eqn:MeanNoONOFF}\\
V_{R}&=& \lambda \mathds{E}[e_n]\mathds{E}[L_n]\mathds{E}[G_n]. \label{eqn:VarNoONOFF}
\end{eqnarray}
Equations~\ref{eqn:MeanNoONOFF} and \ref{eqn:VarNoONOFF} give the mean
and variance of the aggregate data rate of video streaming traffic
when the strategy of no ON-OFF cycles is used to stream videos.

We now show that when users do not interrupt the video download,
\emph{the mean and variance of the data rate are independent of the
streaming strategy used}. Let $D_n'(>D_n)$ denote the time required to
download the video when the video contents are downloaded using either
the short ON-OFF cycles or the long ON-OFF cycles streaming
strategy. For the $n$-th video, the download rate is $G_n$ during the
ON periods and 0 in the OFF periods. If the download rate does not
change during the data transfer, then $\int_{0}^{D_n}X_{n}^{2}(u)du = 
\int_{0}^{D_{n}^{'}}X_{n}^{2}(u)du = e_nL_nG_n$, which leads to the
same variance as in \autoref{eqn:VarNoONOFF}. Using the same argument
and the framework in
Barakat~\etal~\cite{Barakat_2002_FlowModelInternetBackbone}, one can
extend this result to higher moments of the aggregate traffic.   

Therefore, when users do not interrupt the video downloads, we
conclude the following: 
\begin{enumerate}
\item Equations \ref{eqn:MeanNoONOFF} and \ref{eqn:VarNoONOFF} can be
used to dimension the network for video streaming. A simple rule would
be to set the bitrate of links carrying video streaming traffic to
$E[R(t)] + \alpha \sqrt{V_r}$, where $\alpha \ge 1$ is a constraint on
the tolerable bandwidth violations.  
\item The mean and variance of the aggregate data rate of video
streaming traffic are independent of the underlying streaming
strategies used, and hence the required bandwidth. This is important
as video services, where the users are expected to view the whole
video and not interrupt the video download, can safely select a
streaming strategy that can be optimized for other goals such as
server load without overwhelming the network. 
\item An increase in the video encoding rate, for example when YouTube
increases the default video resolution, shall increase the aggregate
rate of video traffic. However, because the variance is a
\emph{linear} function of the video encoding rate, the aggregate
traffic shall be \emph{smoother} than the aggregate traffic observed
at lower encoding rates.  
\end{enumerate}

\begin{table}
\centering
\begin{small}
\begin{tabular}{|c|m{0.8\columnwidth}|}
\hline
Name & Description\\
\hline
$\lambda$ & Arrival rate of videos streaming sessions.\\
\hline
$n$ & number of videos.\\
\hline
$e_n$ & Encoding rate of the $n$-th video.\\
\hline
$L_n$ & Duration (or length) of the $n$-th video.\\
\hline
$B_n$ & Buffering amount for the $n$-th video.\\
\hline
$B'_n$ & Buffering amount for the $n$-th video in terms of playback time.\\
\hline
$S_n$ & Size of the $n$-th video $S_n = e_nL_n$.\\
\hline
$k_n$ & The accumulation ratio for the $n$-th video.\\
\hline
$\beta_n$ & Users interrupt the $n$-th video after time $\beta_n L_n$.\\
\hline
$R(t)$ & Aggregate data rate of streaming traffic at time $t$.\\
\hline
$R'(t)$ & Aggregate amount of bandwidth wasted at time $t$ when users
interrupt video download due to lack of interest.\\
\hline
\end{tabular}
\end{small}
\caption{Variables used in the model.}
\label{tab:ModelVariables}
\end{table}

\subsection{Video Download with Interruptions}
\label{sec:ModelInterrupt}

Users can interrupt a streaming session due to various reasons such as
poor playback quality or lack of interest in the given video. When a
user interrupts the video download due to lack of interest, the data
downloaded but not used by the player is wasted. The wastage of
network resources can be quantified using the amount of unused
bytes. The amount of unused bytes due to lack of interest is important
because Gill~\etal~\cite{Gill_2007_Youtube} observe that 80\% of the
video interruptions in a campus network are due to lack of user
interest. According to
Finamore~\etal~\cite{Finamore_2011_YouTubeDeviceInfra}, 60\% of the
YouTube videos are watched for less than 20\% of their
durations. Similarly, Huang~\etal~\cite{Huang_2007_VODProfit} show
that viewing time decreases as the duration of the video increases.

We now present the impact of the buffering amount and the accumulation
ratio on the amount of unused bytes. We assume that the user
interrupts the download of the $n$-th video after time $\tau_n$ from
the start of the video playback. We further assume that the amount
downloaded in the buffering phase is $B_n$, $B_n \ge 0$, and the time
required for downloading this amount is negligible. If $G_n$ is the
average download rate in the steady state phase, then the amount of
data that can be downloaded up to time $\tau_n$ is $B_n +
G_n\tau_n$. We keep denoting the encoding rate and duration of the
$n$-th video as $e_n$ and $L_n$ respectively. Thus, the interruption
of the $n$-th video shall take place before the whole video has been
downloaded only if  
\begin{equation}
e_nL_n > B_n + G_n\tau_n \ge e_n\tau_n.  
\label{eqn:InterruptCriteria}
\end{equation}

We now assume the download rate of the $n$-th video is
limited by the accumulation ratio $k_n = \dfrac{G_n}{e_n}$, $k_n \ge
1$. We also assume that $\tau_n = \beta_n L_n$, where $\beta_n$, 
$\beta_n < 1$, is the fraction of the $n$-th video watched before 
interruption. \autoref{eqn:InterruptCriteria} can now be written
as 
\begin{equation} 
e_nL_n > B_n + e_nk_n\beta_n\L_n \ge e_n\beta_nL_n.  
\label{eqn:InterruptCriteriaKBeta}
\end{equation}

When $B_n = e_nB'_n$, where $B'_n$ is the amount of playback time buffered
in the buffering phase, the left hand side of
\autoref{eqn:InterruptCriteriaKBeta} can be written as 
\begin{equation} 
B'_n < L_n (1 - k_n \beta_n).
\label{eqn:InterruptCriteriaKBetaPrime}
\end{equation}
In \autoref{sec:YouTubeShort} we observed a buffering of 40~seconds
worth of playback, and an accumulation ratio of 1.25 for Flash
videos. When a user interrupts the video download after watching 20\%
of the video, substituting $B'_n = 40$ seconds, $k_n = 1.25$, and
$\beta=0.2$ yields $L_n = 53.3$ seconds. This implies that, assuming a
fast buffering, YouTube Flash videos that have a duration smaller than
$53.3$ seconds will be downloaded before the viewers have seen $20\%$
of the video. 

We now use the amount of unused bytes to obtain the average bandwidth
wasted due to user interruption. When the $n$-th user 
interrupts the video download at time $\tau_n$, then the amount of
bytes downloaded is $min(B_n+G_n\tau_n, e_nL_n)$. The total amount of
bytes consumed by the player up to time $\tau_n$ is
$e_n\tau_n$. Therefore, the amount of unused bytes is
$min(B_n+G_n\tau_n, e_nL_n) - e_n\tau_n$, and the average bandwidth
wasted is given by  
\begin{equation}
\mathds{E}[R'(t)] = \lambda\mathds{E}[min(B_n+G_n\tau_n, e_nL_n) -e_n\tau_n]. 
\label{eqn:MeanUserIntWasted}
\end{equation}
When the accumulation ratio of the $n$-th video is $k_n$ and the user
interrupts the video after viewing $\beta_n$ fraction of the video,
then substituting $B_n + G_n\tau_n = e_nB'_n + e_nL_nk_n\beta_n$ in
\autoref{eqn:MeanUserIntWasted} yields       
\begin{equation}
\mathds{E}[R'(t)] = \lambda \mathds{E}[e_n]\mathds{E}[min(B'_n +k_n\beta_nL_n, L_n) - \beta_nL_n].
\label{eqn:MeanUserIntWastedAccBeta}
\end{equation}

In summary, \autoref{eqn:InterruptCriteriaKBetaPrime} provides a
condition to limit the amount of unused bytes when users interrupt the
video download due to lack of
interest. Equations \ref{eqn:MeanUserIntWasted} and 
\ref{eqn:MeanUserIntWastedAccBeta} can be used to compute the amount 
of bandwidth wasted due to user interruptions. 

\section{Related Work}
\label{sec:RelatedWork}

Maier~\etal~\cite{Maier_2009_ResBroadband}, and
Labovitz~\etal~\cite{Labovitz_2010_InternetTraffic200Exa}, show that
video streaming contributes to 25-40\% of all HTTP
traffic. Due to its growing popularity, video streaming has received 
considerable attention in the last few years. 

A significant amount of research has been on the video contents
characterization on YouTube and on viewing patterns on
YouTube. Cha~\etal~\cite{Cha_2007_Youtube} study the popularity of
videos and propose caching techniques to enhance the user
experience. Zink~\etal~\cite{Zink_2009_YouTube} study the viewing
patterns in a campus network and suggest proxy caches for enhancing
the user experience and reducing the network
traffic. Gill~\etal~\cite{Gill_2007_Youtube} study the viewing
patterns in a campus network and show that 80\% of user interrupts in
their campus were due to lack of user interest. Similarly,
Finamore~\etal~\cite{Finamore_2011_YouTubeDeviceInfra} show that 60\%
of the YouTube videos are watched for less than 20\% of their
duration.

These works are orthogonal to ours. Indeed, we focus on the network
traffic characterization of YouTube and Netflix, not on the content
characterization. 

Plissonneau~\etal~\cite{Plissonneau_2008_WebYouTube},
Saxena~\etal~\cite{Saxena_2008_VideoWebNossdav}, and
Alcock~\etal~\cite{Alcock_2011_YouTubeFlowControl} observe rate
limitations on YouTube traffic. They do not identify the streaming
strategies discussed in our paper.
Akhshabi~\etal~\cite{Akhshabi_2011_StreamingRateAdaption} only
observed a rate limitation in the steady state phase for
Netflix. Saxena~\etal~\cite{Saxena_2008_VideoWebNossdav} show that the
YouTube videos streamed using the servers of Google have a buffering
phase, whereas the legacy servers of YouTube do not show this
buffering phase. Alcock~\cite{Alcock_2011_YouTubeFlowControl} only
characterized the strategy for short ON-OFF cycles for Flash videos on
YouTube.

To the best of our knowledge, we are the first to identify and
characterize the three streaming strategies used by YouTube and
Netflix. Moreover, we derive a mathematical model to study the
aggregate traffic due to video streaming. Therefore, our work enhances
significantly previous knowledge on video streaming.   

\section{Conclusion}
\label{sec:Conclusion}

In this paper, we present an in depth traffic characterization of
Netflix and YouTube. We identify three different streaming
strategies with fundamentally different traffic properties. We show
that Netflix and YouTube adapt the streaming strategy depending on the
application and the container used. This is a concern as it
means that a sudden change of application or container in a large
population might have a significant impact on the network traffic. 
Considering the very fast changes in trends this is a real
possibility, the most likely being a change from Flash to HTML5
along with an increase in the use of mobile devices.  

We derive a model for the aggregate traffic generated by the
different streaming strategies. We use this model to show that streaming
videos at high resolutions can result in smoother aggregate traffic
while at the same time linearly increase the aggregate data rate due
to video streaming. We also show how the amount buffered and the
accumulation ratio can be adapted considering the user interruptions
due to lack of interest.  

However, we did not consider the impact of the three different
streaming strategies on the network loss rate. We believe that it will
have less impact than the wasted bandwidth due to lack of users
interest that we studied in this paper. It is anyway a possible area
of improvement.

\section{Acknowledgment} 

The research leading to these results has received funding from the 
The French National Research Agency (ANR) - Connect project
(www.anr-connect.org). 

This work was supported by the ARO under grant MURI W911NF-08-1-0233.
The views and conclusions contained in this document are those of the
authors and should not be interpreted as representing the official
policies, either expressed or implied of the ARO.

\end{document}